\documentclass[graybox]{svmult}

\usepackage{type1cm}        
\usepackage{makeidx}         
\usepackage{graphicx}        
\usepackage{multicol}        
\usepackage[bottom]{footmisc}
\usepackage{placeins}
\usepackage[numbers]{natbib}
\usepackage{multirow}
\usepackage{newtxtext}       %
\usepackage{newtxmath}       
\usepackage{float}       
\usepackage{subcaption}
\usepackage[super]{nth}
\usepackage{graphicx} 
\usepackage{hyperref}

\makeindex             

\begin{document}


\title*{Computational study on an Ahmed Body equipped with simplified underbody diffuser}
\author{Filipe F. Buscariolo, Gustavo R. S. Assi and Spencer J. Sherwin}
\institute{Filipe F. Buscariolo \at Imperial College London, NDF-USP, McLaren Racing \email{f.fabian-buscariolo16@imperial.ac.uk}
\and Gustavo R. S. Assi \at NDF-USP \email{g.assi@usp.br}\and Spencer J. Sherwin \at Imperial College London \email{s.sherwin@imperial.ac.uk}}
%
%
\maketitle


\abstract{The Ahmed body is one of the most studied 3D automotive bluff bodies and the variation of its slant angle of the rear upper surface generates different flow behaviours, similar to a standard road vehicles. In this study we extend the geometrical variation to evaluate the influence of a rear underbody diffuser which are commonly applied in high performance and race cars to improve downforce. Parametric studies are performed on the rear diffuser angle of two baseline configurations of the Ahmed body: the first with a 0$^\circ$ upper slant angle and the second with a 25$^\circ$ slant angle. We employ a high-fidelity CFD simulation based on the spectral/hp element discretisation that combines classical mesh refinement with polynomial expansions in order to achieve both geometrical refinement and better accuracy. The diffuser length was fixed to the same length of 222 mm similar to the top slant angle that have previously been studies. The diffuser angle was changed from 0$^\circ$ to 50$^\circ$ in increments of 10$^\circ$ with an additional case considering the angle of 5$^\circ$. The proposed methodology was validated on the classical Ahmed body considering 25$^\circ$ slant angle, found a difference for drag and lift coefficients of 13\% and 1\%, respectively. For the case of an 0$^\circ$ slant angle on the upper surface the peak values for drag and negative lift (downforce) coefficient were achieved with a 30$^\circ$ diffuser angle, where the flow is fully attached with two streamwise vortical structures, analogous to results obtained from \cite{ahmed1984some} but with the body flipped upside down. For diffuser angles above 30$^\circ$, flow is fully separated from the diffuser. The Ahmed body with 25$^\circ$ slant angle and a diffuser achieves a  peak value for downforce at a 20$^\circ$ diffuser angle, where the flow on the diffuser has two streamwise vortices combined with some flow separation. The peak drag value for this case is at 30$^\circ$ diffuser angle, where the flow becomes fully separated.
}

\section{Introduction}
\label{sec:intro}

Among the standard automotive bluff bodies in literature, the most studied one is the Ahmed body, firstly proposed by Ahmed \cite{ahmed1984some}. It is based on the geometry designed by Morel \cite{morel1978aerodynamic}, with the main dimensions highlighted in Figure~\ref{fig:Ahmed_dim}. The proposed geometry of the Ahmed body aims to reproduce the main features of road vehicles, such as the frontal stagnation, ground effect and well-defined separation points. 

\begin{figure}[b]
\begin{subfigure}[b]{1\textwidth}
\begin{center}
\includegraphics[width=1\textwidth]{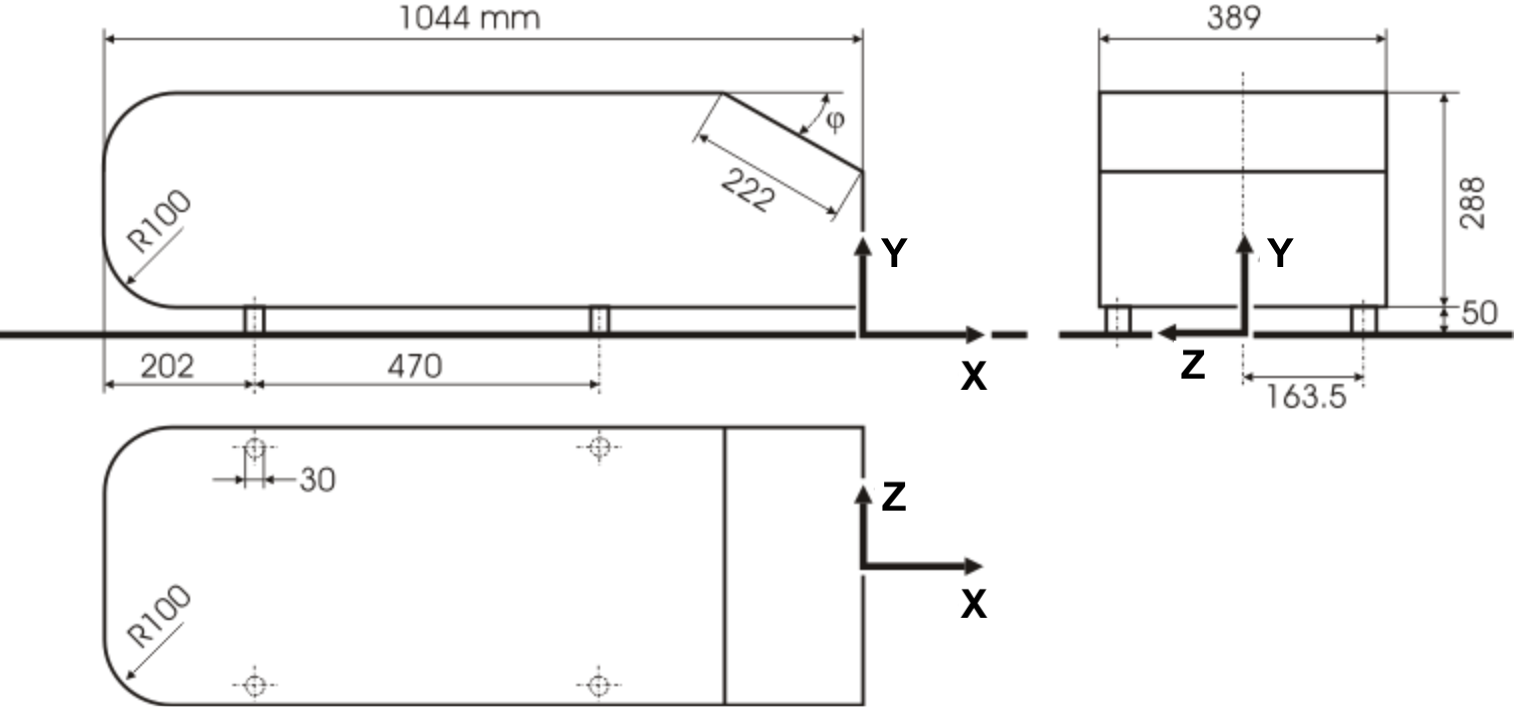}
\label{fig:Ahmed_dim}
\end{center}
\end{subfigure}
\begin{subfigure}[b]{1\textwidth}
\begin{center}
\includegraphics[width=0.8\textwidth]{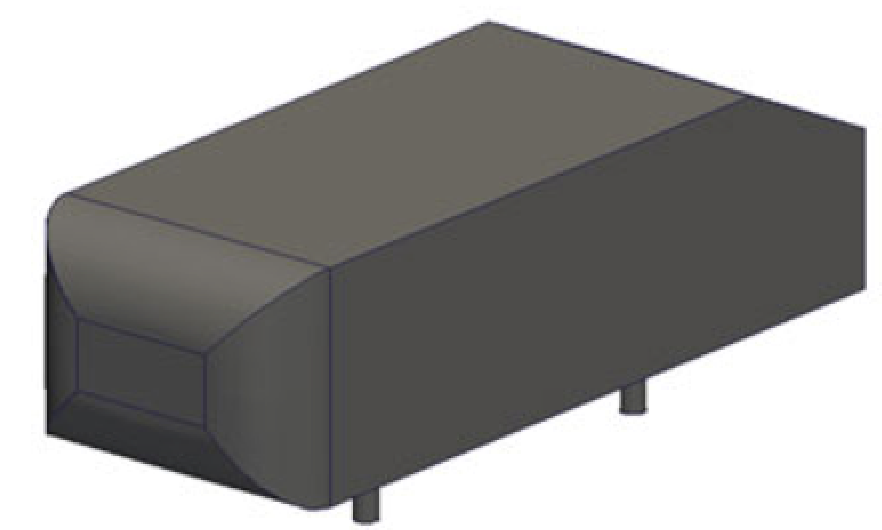}
\label{fig:Ahmed_3d}
\end{center}
\end{subfigure}
\caption{Ahmed body schematic drawing. The upper slant length of 222 mm is fixed, independent of its inclination angle $\varphi$.}
\label{fig:Ahmed_dim}
\end{figure}

The most emblematic characteristic of the Ahmed Body is a angled upper back section with fixed length, here referred as slant, on the upper rear portion, allowing the simulation of different automotive body styles. According to (\cite{huminic2010computational}), it has been shown that the flow over the slanted surface back section is dependent on specific inclination angle. Two critical angles, at 12.5$^{\circ}$ and 30$^{\circ}$ have been observed, in which the flow structure changes significantly, and where a change of curvature of the drag coefficient is also evident. For angles below 12.5$^{\circ}$, the airflow over the slant remains fully attached before separating from the model when it reaches the rear of the body. The flow from the angled section and the side walls produces a pair of counter rotating vortices, which then persist downstream. 

For angles between 12.5$^{\circ}$ and 30$^{\circ}$, the flow over the slant becomes highly complex. Two counter-rotating lateral vortices are shed from the sides of the angled section with increased size, which affects the flow over the whole back end, specially the previously existing three-dimensional wake. These vortices are also responsible for maintaining attached flow over the rest of the angled surface up to a slant angle of 30$^{\circ}$, and it has been shown that they are extended up to half of the length of the model beyond the trailing edge, as discussed in \cite{strachan2007vortex} and \cite{lienhart2003flow}. Close to the second critical angle, a separation bubble is also formed over the inclined slant. The flow separates from the body, but re-attaches before reaching the vertical back section. 

Above a 30$^{\circ}$ slant angle, the flow over the or slant is fully separated. However there is a weak tendency of the flow to turn around the side of the model, as a result of the relative separation positions of the flow over model top and those over the slant side edges. When the flow is in this state, a near constant pressure is found across that region. To characterize all three flow configurations here discussed, representative slant angles are commonly used in literature with 0$^{\circ}$ (or squared-backed), 25$^{\circ}$ and 35$^{\circ}$y.


The first experimental study on Ahmed Body \cite{ahmed1984some} was with static floor conditions, considering a Reynolds number $Re = 4.29 \times 10^{6}$ based on its full length. In this study, results for the drag coefficient were obtained for different slant angles, ranging from 0$^{\circ}$ to 40$^{\circ}$, in increments of 5$^{\circ}$ with an additional measurement at 12.5$^{\circ}$. Due to limitations on the wind tunnel setup, only drag force measurements and a few flow visualization test were performed. In order to better understand the flow phenomena and turbulence structures around the model, a complementary study was performed by \cite{lienhart2003flow} using laser Doppler anemometry (LDA), hot-wire anemometry (HWA) and static pressure measurements. 

\begin{figure}[hbt]
	\begin{center}
	\includegraphics[width=0.6\textwidth]{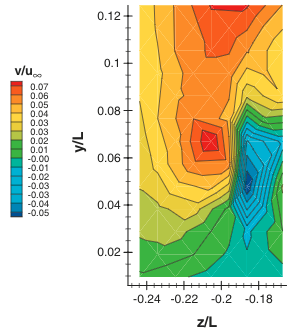}
	\caption{Normalized V velocity on a squared-back Ahmed Body at x/L = 0.048. Reproduction from \cite{strachan2007vortex}.}
	\label{fig:lower_vortex_0}
	\end{center}
\end{figure}

Aiming to reproduce realistic road conditions and understand the phenomena associated to flow fields close to the ground, the authors of \cite{strachan2007vortex} performed an Ahmed Body wind tunnel test using moving ground and acquired both the aerodynamic forces and the flow characteristics by employing time-averaged LDA. The flow conditions were also slightly different from the ones used on Ahmed's first test, since it had a Reynolds number of $Re = 1.7 \times 10^{6}$. Nevertheless, similar flow behaviour were observed on the slant, despite the quantitative results being slightly different. One of the most interesting features found in this flow visualization results is the lower vortex system, a pair of vortices that appears close to the ground interface, which were absent in the fixed-ground studies. According to \cite{strachan2007vortex}, this could be attributed to the interference caused by the four studs used to support the model on the floor. Figure~\ref{fig:lower_vortex_0} illustrates this phenomenon on an Ahmed Body with a squared-back.

An important development in automotive industry directly associated with the flow near the ground is the introduction of underbody diffusers, initially for high performance race vehicles with relatively low ground clearance. By providing a smoother transition from the underbody flow to the base of the car body, the strength of the rear wake can be reduced, contributing to drag reduction. In addition, it was found that at slightly inclined angles, the underbody diffuser also increases the downforce generated, assisting the acceleration and handling.

To explore detailed features of the near-ground vortices, and to examine the potential benefits of implementing underbody diffusers, we propose a series of computational studies considering same simulation conditions as the experiment from \cite{strachan2007vortex}, with moving ground. The Ahmed bodies used in the study are the squared-back and the slant angle of 25$^{\circ}$, representing respectively estate car/station-wagons (attached flow) and performance cars (vortex generation with flow detachment). The length of the underbody diffuser is set to be the same as the classic Ahmed body slant length, with angles ranging from 10$^{\circ}$ to 50$^{\circ}$, in increments of 10$^{\circ}$. An additional case also considers a diffuser angle of 5$^{\circ}$, a setting commonly found in racing vehicles. This study focus essentially on the aerodynamic quantities on the Ahmed body, as well as, on the flow structures on its geometry, such as the vortices on the slant and diffuser.

CFD has become an underpinning technology for most automotive companies to reduce development times and costs. Since the Ahmed Body is a widely studied bluff body, it has become a test case to validate new CFD codes, specially for applications in the automotive industry. Lower vortices observed by \cite{strachan2007vortex} were not present in CFD simulation studies with fixing studs modelled. They were first observed by \cite{krajnovic2004large}, where an Ahmed Body with slant angle of 25$^{\circ}$ was simulated without the fixing studs. Nevertheless, the location where these vortices are generated and possible interactions with underbody components were not highlighted. 

We utilise a high fidelity spectral/hp element method simulation using under-resolved direct numerical simulation (uDNS)  also known as implicit large eddy simulation (iLES)(\cite{article2007iles}).  The spectral/hp elemental method combines the advantages of higher accuracy and rapid convergence from the spectral (p) methods, while maintaining the flexibility  of the classical finite element (h) complex meshes, allowing unsteady vortical flows around geometries to be effectively captured. We present the validation of the proposed numerical methodology on the classical Ahmed body with 25$^{\circ}$ slant angle, as in the study of \cite{filipe2018ahmed}. The Ahmed body with 25$^{\circ}$ slant angle, although in the pre-critical regime, still poses a challenge for most CFD codes due to the complex flow physics, however, it is a well-established test configuration, as performed by \cite{lienhart2003flow} and  \cite{strachan2007vortex}.

Most computational studies on the Ahmed Body employ simplified Reynolds Averaged Navier-Stokes (RANS) solution. This approach is very reliable for simple stable flow problems, however it is not suitable to correctly predict unstable phenomena around complex geometries. In the study by \cite{krajnovic2004large}, for the first time for an Ahmed Body, a LES methodology was used yielding solutions of higher flow details, especially for the critical slant angle of 25$^{\circ}$. A major limitation of running LES or detached eddy simulation (DES) for this kind of geometry is the requirement of high mesh resolution, with considerably higher simulation cost and time. 

The latest achievements in the high-fidelity turbulence models around an Ahmed body are found mainly for the slant angle of 25$^{\circ}$ and are summarized in the compilation work of \cite{serre2013simulating} in which a comparative analysis of recent simulations, conducted in the framework of a French-German collaboration on LES of Complex Flows at Reynolds number of 768,000. The study offers a comparison between results obtained with different eddy-resolving modelling approaches: three classical h-type method (LES with Smagorinsky model and wall function (LES-NWM), Wall-resolving LES with dynamic Smagorinsky model (LES-NWR), and DES with shear stress tensor (DES-SST)) and one spectral element method (implicit LES with spectral vanishing viscosity (iLES-SVV)). The iLES-SVV simulation in \cite{serre2013simulating} work was conducted in various two-dimensional planes along the span-wise direction, and subsequently constructed into three-dimensional flow fields (commonly known as 2.5D simulation). Considering the drag coefficient, both LES-NWR and DES-SST overestimated the value in around 16\%; the LES-NWM presented a difference around 6\%, which presented the best agreement. The iLES-SVV model better modelled the flow behaviour compared to previous models, however the drag difference was around 44\%

A new Improved Delayed DES (IDDES) methodology, an enhancement of the Delayed DES (DDES), is proposed by \cite{guilmineau2017assessment} to solve the flow around the Ahmed body. The study presents a comparison between quantitative and qualitative results obtained with different methodologies previously presented with this newly proposed methodology. The IDDES case is the one that most closely correlates the flow behaviour and structures with experimental reference. However, results of the aerodynamic quantities are different, such as the drag coefficient with approximately 27\% difference from same experimental reference.

\section{Ahmed body equipped with rear underbody diffuser}
\label{sec:ahmed_dif}

Bluff bodies equipped with rear underbody diffusers are being studied by several researchers, especially from the automotive industry, to maximise the performance of the vehicle. The study of \cite{cooper1998aerodynamic} identified three important characteristics on a body underbody diffuser. The first is a diffuser pumping effect, which occurs once the outlet of the diffuser is set as the base pressure of the body, as identified by \cite{jowsey2013experimental}.  The diffuser recovers pressure along its length, considering continuity and applying an inviscid, steady argument of constant total pressure using Bernoulli's equation implies that the diffuser inlet pressure should be reduced, causing a suction effect. The second characteristic is the interaction with the road, in which as the ground clearance between the floor and the underbody becomes smaller, flow velocity in that region increases and pressure drops, following the same continuity and Bernoulli's equation. The third characteristic is the angled upsweep, which generates vortices on the diffuser up to a certain critical angle, creating an upwash of the flow, aiding flow attachment and increasing downforce. 


Complementing the work of \cite{cooper1998aerodynamic}, \cite{senior2001force} investigated a new bluff body equipped with a diffuser  which extended over 41\% of the body length and  with inclination angle of 17$^\circ$ and endplates in different ground heights. The result was the identification of four distinct regions of diffuser performance, all related to the model ground height. The first region from non-dimensional ride height $h/H$, where $h$ is the distance from the body to the ground and $H$ is the total height of the body, is defined from 0.76 to 0.38 and is defined as downforce enhancement, region where the flow on the diffuser is symmetric with some separation on the diffuser inlet. The second region, referred as maximum downforce, from $h/H$ 0.38 to 0.22, with similar flow behaviour as the first region, except for the formation of a separation bubble at the center of the diffuser. The third and fourth regions are referred both as the downforce reduction from $h/H$ 0.22 and low downforce region from $h/H$ 0.16. The third region is characterized by a sudden drop of downforce performance and the fourth region shows that further ground height reduction causes the downforce to be reduced and this fact is explained by an asymmetric and separated flow behaviour at the diffuser inlet.

With substantial literature on the diffusers, such as the works from \cite{cooper1998aerodynamic} and \cite{senior2001force}, we notice that each study has employed different bluff body geometries, that have not been previously studied without a diffuser. The work of \cite{huminic2010computational} was the first to propose a computational study of diffusers on an Ahmed body with slant angle of 35$^\circ$ with similar conditions of \cite{strachan2007vortex} study. This body style has a quasi-2D behaviour and a combination of five diffuser lengths with eight different angles were evaluated to predict drag and lift coefficients. The flow physics was not fully evaluated and no conclusions were found mainly due to the nature of the averaged flow solution employed on the study. Following the previous work, \cite{huminic2012numerical} present a similar study, considering the Ahmed body with slant angle of 35$^{\circ}$ with a rear underbody diffuser, wheels and wheelhouses.

The work of \cite{moghimi2018numerical} performed an experimental and numerical study on an Ahmed body equipped with diffuser. Both experiments and simulations were performed at a low Reynolds number of $Re = 9.31 \times 10^4$ considering a 10\% scale half Ahmed body with slant angle of 25$^{\circ}$. This study offers an experimental reference, however due to the model scale factor being extremely reduced compared to other experiments, the geometry becomes more sensitive to surface imperfections, introducing an additional source of error in the measurements. The Reynolds number is also reduced compared to other Ahmed body experimental references.

A common point on the literature presented considering the Ahmed body with diffusers is the use of CFD simulations. The CFD simulations for all references presented employ the RANS methodology combined with the $k-\omega SST$ turbulence model and no reference using high-fidelity CFD solutions on the Ahmed body with diffuser were previously presented.

The geometry of the diffuser is proposed without the use of endplates. By evaluating the Ahmed Body equipped with a rear underbody diffuser without endplates, we also offer an interesting and simple test case, especially for the squared-back case, as it can be evaluated using a regular Ahmed body which has been flipped upside-down.

\section{uDNS/iLES simulations using spectral/hp element method}
\label{sec:spectral}

\begin{figure}[hbt]
	\begin{center}
	\includegraphics[width=1\textwidth]{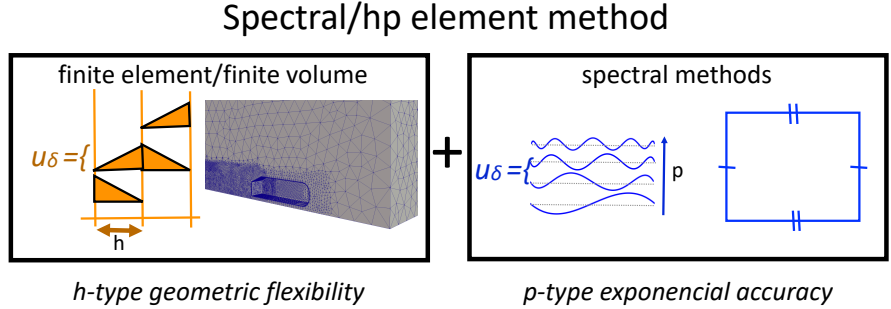}
	\caption{Schematic explanation illustrating how finite element (h) and spectral methods (p) combine to form the spectral/hp element method.}
	\label{fig:elements}
	\end{center}
\end{figure}

For both Ahmed body styles with diffuser, we performed implicit LES simulations based on a spectral/hp element approach. Classical h-type method is based on dividing the domain into non-overlapping elements of the same type, similar as in the Finite Element Method (FEM), offering geometric flexibility, a key factor for many complex industrial cases. To improve the accuracy of the solution, the mesh characteristic length (h) is reduced in order to capture smaller flow features, generating a finer mesh, with larger element density of the same type. The p-type method focus on improving results by increasing the degree of the polynomial expansion used to approximate the solution on each element on a fixed mesh, with the desirable property of exponential convergence. The spectral/hp element method used in this work combine both spatial approximations (h and p) in order to have a methodology flexible enough to handle complex geometries and providing high-fidelity solutions, such as LES and DNS with enhanced convergence properties. A summary of the methods is illustrated in Figure\ref{fig:elements}.

The flow solution for the cases in this work is obtained by using the incompressible Navier-Stokes solver with a velocity correction scheme as proposed by \cite{guermond2003velocity}. The elliptic operators were discretised using a classical continuous Galerkin (CG) formulation and all this formulation is encapsulated in the open source package Nektar++ \cite{cantwell2015nektar}.

In this work we adopt an equivalent of the Taylor Hood approximation, approximating the velocity by continuous piecewise quadratic functions and the pressure by continuous piecewise linear functions.Therefore we use one higher polynomial order for velocity than the pressure.  The polynomial order for velocity is also referred to as the simulation expansion order in this study.

For simulations with higher Reynolds numbers ($10^{5}$ and above), such as the cases presented here, the flow is typically only marginally resolved, which means that the ratio of subgrid scale (SGS) and resolved dissipation is relatively small. Such a marginal resolution can lead to numerical instabilities related to wave interaction and wave trapping. To reach a stable solution, we employ both dealiasing and spectral vanish viscosity (SVV) stabilization techniques.

The aliasing errors related to the Navier-Stokes equations appears when handling its quadratic non-linearity term by using the Gauss integration orders $Q$ similar to the solution polynomial order $P$. This is usually present in simulations considering under-resolved turbulence, such the iLES, which leads to a significant error increment in high-frequency modes of the solution and typically cause the simulations to diverge. To avoid the aliasing errors, we employed a quadrature order consistent with the polynomial order and non-linearities of the equation. In areas of non-linear geometry deformation we also have to be mindful of geometric aliasing (aliasing arising from geometric mapping). We refer the interested reader to \cite{mengaldo2015dealiasing} for more details. 

High-order methods present low numerical diffusion, as discussed by \cite{karniadakis2013spectral} and simulations with marginal spatial resolution become numerically unstable, especially in the presence of turbulence at high Reynolds numbers, condition in which most of engineering problems are. The spectral-vanishing viscosity (SVV) is a technique that adds artificial dissipation to the smallest scales of the solution and this modern strategy has been proven to effectively stabilise the simulation.

The main idea of SVV consists in expanding the Navier-Stokes Equations to include an artificial dissipation operator, leading to:

\begin{subequations}
\begin{align}
\label{eqn: SVV1}
\frac{\partial \boldsymbol{u}}{\partial t} & = - (\boldsymbol{u}	\cdot \nabla) \boldsymbol{u} -\nabla p + \nu \nabla^2 \boldsymbol{u} + S_{V V} (\boldsymbol{u})\\
\nabla \cdot \boldsymbol{u} & = 0
\end{align}
\end{subequations}

and the original operator SVV is:

\begin{equation}
\label{eqn: SVV2}
S_{V V} (\boldsymbol{u}) = \epsilon \sum_{i = 1}^{N_{dim}} \frac{\partial}{\partial x_i} [Q_i * \frac{\partial u}{\partial x_i}]
\end{equation}

with ${N_{dim}}$ being the spatial dimension of the problem, $\epsilon$ a constant coefficient, and $*$ representing the application of the filter $Q_i$ through a convolution operation.

For the SVV operator in this study, we run the simulation using a novel CG-SVV scheme with DGKernel as proposed by \cite{moura2017eddy}. The fundamental idea of this implementation is based on the fixing of the P\'eclet number, which can be understood as a numerical Reynolds number based on local velocity and mesh spacing. This is achieved by making the viscosity coefficient of the SVV operator proportional to both a representative velocity and a local mesh spacing. Once the P\'eclet number is the same for the domain, the authors in \cite{moura2017eddy} proposed a SVV kernel operator for CG methods that mimics the properties of DG-based discretisation where there is natural damping of high frequency and reflected waves. In this approach the dissipation curves arising from spatial eigenanalysis of CG of order $P_N$ are matched to those of DG with order ${P_N}-2$. Matching both curves offers benefits such as the numerical stability of simulations at very high Reynolds number.

\section{Numerical methodology validation on the classical Ahmed body geometry}
\label{sec:validation}

In this simulation study, we use a coordinate system with $X$ as the streamwise direction, $Y$ as the vertical direction and $Z$ as the spanwise direction. The Ahmed body model is positioned with its back end on the coordinate $X = 0$ and at a distance $h$ of 50 mm from the ground ($Y = 0$). The wind tunnel test section size is defined with same dimensions as the experiment from\cite{strachan2007vortex} 1660 mm $\times$ 2740 mm, keeping the same blockage ratio. Air flow inlet is positioned at $X = - 2 L$ and outlet at $X = 2 L$ with a total $X$ length of $4 L$. With such reduced wind tunnel, special outflow boundary condition is necessary to avoid interference on the flow and numerical instabilities. The outflow condition selected is the high-order outflow condition, proposed by \cite{dong2014robust}, in order to avoid wave reflections back to the domain.

The Reynolds number for all simulated cases is $Re = 1.7 \times 10^{6}$, based on the Ahmed body total length $L$ of 1044 mm. With this Reynolds number value and by imposing moving ground condition, we aim to reproduce similar conditions employed by \cite{strachan2007vortex}.

The high-order meshes for all cases presented in this work were generated by the mesh generator module of Nektar++: NekMesh (\cite{turner2017high}). The pipeline to create a high-order mesh starts by designing the geometries for the computational simulations on a Computer Aided Design (CAD) software, exported in STEP format. Subsequently, we generate a linear mesh using a classical h-type method. Linear mesh generation on NekMesh also incorporates an optimisation step to avoid irregular and low quality elements once the surface mesh is projected into its 3D form using the CAD surface.

Once the linear mesh was generated, the next step is to convert it into a high-order mesh which is geometry conforming. The generation of the high-order mesh requires the addition of extra points to represent the polynomial discretization (with order $P_M$), which are added along the curved edges, CAD surface geometry and in the interior of the domain. The processes then follows by the generation of a macro boundary layer on user-defined wall surfaces together with volumetric high-order mesh on rest the domain. The final operation on the mesh generation is the macro boundary layer split, by using the isoparametric approach as proposed by the authors of \cite{moxey2015isoparametric} considering boundary layer inputted parameters.

\begin{figure}[hbt]
	\begin{center}
	\includegraphics[width=1\textwidth]{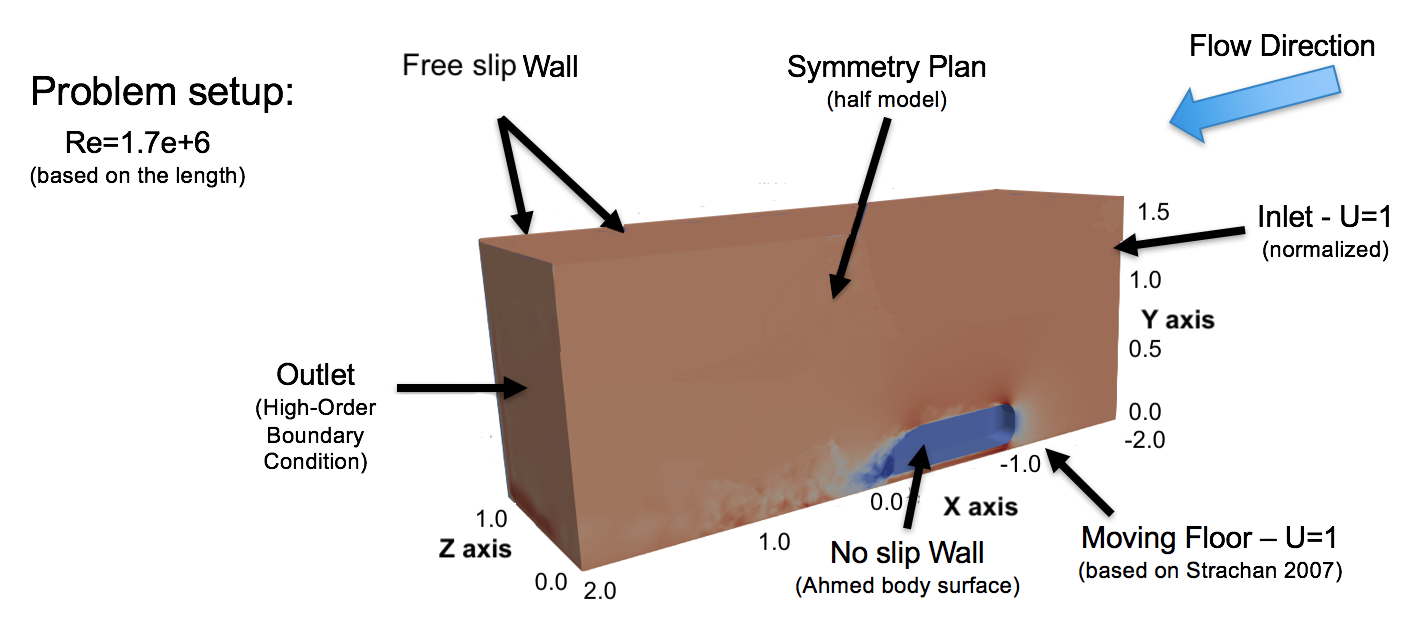}
	\caption{Boundary condition setup.}
	\label{fig:bc}
	\end{center}
\end{figure}

The boundary conditions for the computational study were set as follows and shown on Figure~\ref{fig:bc}:
\begin{itemize}
\item Ahmed bodies with diffuser are set as wall with no-slip condition;
\item A half model of the geometry is used;
\item Symmetry condition imposed at $Z = 0$;
\item Free-slip condition imposed at tunnel walls;
\item Uniform velocity profile at the inlet;
\item High order outflow condition at the outlet (as proposed by \cite{dong2014robust});
\item A moving ground condition on the floor with speed $U$ in the $X$ direction, as used by \cite{strachan2007vortex};
\item Convergence criteria for pressure is $1 \times 10^{-6}$ and for the velocity components $1 \times 10^{-8}$;
\end{itemize}

Simulations were performed for 7 convective time units (CTU), which can be understand as the free stream flow that has been advected over a surface or reference point for a total length of $7 L$. The convergence criteria was set to a maximum variation of $1 \times 10^{-6}$ for pressure and $1 \times 10^{-8}$, in order to minimize numerical error. Due to the use of half symmetric model, this study focus mainly on aerodynamic quantities, such as the lift and drag coefficients and flow structures on the slant and close to the body. The use of half symmetric Ahmed body was previously validated against full geometry by \cite{buscariolo2019spectral}. Under similar boundary conditions, solution and meshing configurations, the maximum difference in the aerodynamic quantities was 2\%. 

Considering the numerical methodology validation, as performed by \cite{filipe2018ahmed}, we are focusing on the classical Ahmed body geometry, selecting the slant angle of of 25$^{\circ}$. Two different mesh configurations in terms of h-refinements are proposed, defined as \emph{Original} and \emph{Refined} meshes, based on the  boundary layer setup of a total length of $0.022 L$ (macro boundary layer), which is further divided into 10 layers with a growth rate of 1.6. The possibility of evaluating two different meshes in terms of elements resolution complements the polynomial order $P_N$. The choice of parameters for refinements zones and boundary layer was based on automotive industry guidelines. For the mesh resolution, the study aimed to have similar number of DOFs of a Wall-Modelled LES simulation for the most refined case and further coarsening the mesh to evaluate its impact on the aerodynamic behaviour of the body.

\begin{figure}[hbt]
	\begin{center}
	\includegraphics[width=1.0\textwidth]{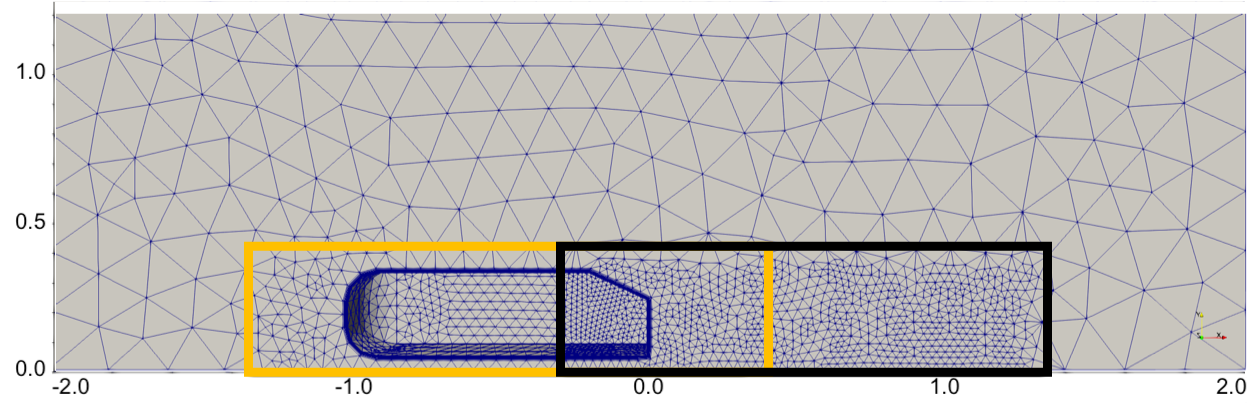}
	\caption{Plane view $Z=0$ indicating the location of the refinement boxes on the Ahmed body model for the \emph{Original} mesh case. The Ahmed body refinement region is indicated in yellow and the wake refinement region is indicated in black.}
	\label{fig: mesh1ahmed}
	\end{center}
\end{figure}

This validation study also proposes the evaluation of three high-order meshes for both \emph{Original} and \emph{Refined} cases, considering the first $P_M$ of \nth{4} order, as previous reference studies from \cite{turner2017framework}. In addition to previous reference, we propose increasing the polynomial resolution of the curved elements $P_M$ to both \nth{5} and \nth{6} order. Within this, the study aims to evaluate the influence of different high-order polynomials for the boundary layer mesh element curvature, when using relatively coarse meshes. Two refinement zones are applied for all Ahmed body simulations. The first refinement zone is created over the Ahmed Body full length, referred as Ahmed body refinement, ranging from 0.3 $L$ before and 0.3 $L$ after the end of the body. The second refinement, referred as wake refinement is applied on the wake region, intercepting the first region in -0.3 $L$ before the end of the body, to 1.3 $L$ after the end of the body. The wake refinement and overlap with smaller elements are applied in order to fully capture the flow phenomena in the separation region. Mesh refinement regions considering the \emph{Original} mesh are illustrated in Figure~\ref{fig: mesh1ahmed} on the plane $Z = 0$.

The \emph{Original} mesh is generated with a total number of elements $N_{EL}$ for half model around 94,000 h-type mesh elements, where around 14,000 are prismatic elements $N_P$ and 80,000 are tetrahedra elements $N_T$. The mesh configuration parameters on NekMesh, typically found in the automotive industry, are presented below in function of $L$.

\begin{itemize}
\item Min length $h_{min}$ = 0.01 $L$;
\item Max length $h_{max}$ = 0.2 $L$;
\item Ahmed Body refinement zone length = 0.05 $L$;
\item Wake refinement zone length = 0.03 $L$;
\end{itemize}

For the \emph{Refined} mesh, the number of elements generated $N_{EL}$ for half model is around 335,000 h-type mesh elements, where around 35,000 are prismatic elements $N_P$ and 300,000 are tetrahedra elements $N_T$. Mesh parameters are presented as follow:

\begin{itemize}
\item Min length $h_{min}$ = 0.0075 $L$;
\item Max length $h_{max}$ = 0.2 $L$;
\item Ahmed Body refinement zone length = 0.035 $L$;
\item Wake refinement zone length = 0.02 $L$;
\end{itemize}

Once the six high-order meshes are generated, high fidelity numerical simulations are performed, using the implicit LES simulations based on a spectral/hp element method. The selected is the Nektar++ open source software for all simulations. The solution on each element is approximated by a high-order polynomial and three different polynomial expansions $P_N$ are selected: $P_N$ = \nth{4}, \nth{5} and \nth{6} order. These are applied for each of the six meshes, in total eighteen cases evaluated. The total number of degrees of freedom (DOFs) of each case, in order to have a comparison with low-order solutions, is presented in Table~\ref{tableDOF}, with numbers indicating in millions of DOFs and (-A) refers to \emph{Original} mesh and (-R) to \emph{Refined} mesh.

\begin{table}[bt]
\begin{center}
\caption{Resolution of the proposed simulations, showing total number of DOF (in Million) for each case evaluated, at different polynomial expansions accuracy $P_N$.}
\label{tableDOF}
\begin{tabular}{|c|c|c|c|c|}
\hline
 & & $P_N=4$ & $P_N=5$ & $P_N=6$ \\ \hline
\multirow{3}{*}{Original Mesh (-A)}  & $P_M=4$ & 2.16 & 3.85 & 6.24 \\ \cline{2-5}
& $P_M=5$ & 2.16 & 3.85 & 6.24 \\ \cline{2-5}
 & $P_M=6$ & 2.16 & 3.85 & 6.24 \\ \hline
\multirow{3}{*}{Refined Mesh (-R)} & $P_M=4$ & 7.40 & 13.13 & 21.21 \\ \cline{2-5}
 & $P_M=5$ & 7.40 & 13.13 & 21.21 \\ \cline{2-5}
 & $P_M=6$ & 7.40 & 13.13 & 21.21 \\ \hline
\end{tabular}
\end{center}
\end{table} 

An important aspect to mention is that the mesh order $P_M$ and solution order $P_N$ are independent and values of both can be freely combined. Higher order of the $P_M$ polynomial are used to improve the reproduction of complex surfaces by curving the elements, whereas $P_N$ adds more DOFs to the solution.

\subsection{Ahmed body drag and lift coefficient validation results}
\label{subsec:AhmedBodyDragandLiftCoefficientCorrelation}

We now compile quantitative results for the nine combinations of high-order meshes with polynomial expansions for the \emph{Original} mesh (-A) and additional nine cases for the \emph{Refined} mesh (-R) in total eighteen cases. The first number following the term NM refers to $P_N$ and the second number refers to the $P_M$ employed to the case. A summary is presented on Table~\ref{tab: draglift1} for the Root Mean Square (RMS) and Root Mean Square Error (RMSE) drag and lift coefficients obtained in all simulations. For the coefficients of drag and lift, results were averaged from the \nth{5} to the \nth{7} CTU, as the results using $P_N = 4$ start to have a more stable profile.

\begin{table}[hbt]
	\begin{center}
	\caption{Drag and lift averaged RMS and RMSE coefficients for the \emph{Original} and \emph{Refined} meshes considering evaluated $P_M$ and $P_N$ high-order, comparing with experiments from \cite{strachan2007vortex}.}
	\label{tab: draglift1}
		\begin{tabular}{|c|c|c|c|c|c|c|}
		\hline \multirow{2}{*}{\textbf{Simulation Case}} & \textbf{$C_D$} & \textbf{$C_L$} & \textbf{Difference} & \textbf{Difference} & \textbf{$C_D$} & \textbf{$C_L$}\\ 
	    & \textbf{RMS} & \textbf{RMS} & \textbf{Drag \%} & \textbf{Lift \%}  & \textbf{RMSE} & \textbf{RMSE}\\
		\hline	NM44-A	&	0.416	&	0.140	&	40	&	-50 & 0.1202 & 0.1637\\
		\hline	NM45-A	&	0.386	&	0.090	&	29	&	-68	& 0.0913 & 0.2384 \\
		\hline	NM46-A	&	0.386	&	0.090	&	29	&	-68	& 0.0912 & 0.2386 \\
		\hline	NM54-A	&	0.322	&	0.270	&	8	&	-3	& 0.0299 & 0.1043 \\
		\hline	NM55-A	&	0.312	&	0.286	&	5	&	2	& 0.0157 & 0.0325 \\
		\hline	NM56-A	&	0.313	&	0.285	&	5	&	2	& 0.0163 & 0.0339\\
		\hline	NM64-A	&	0.276	&	0.267	&	-8	&	-5	& 0.0230 & 0.0544 \\
		\hline	NM65-A	&	0.251	&	0.260	&	-16	&	-7	& 0.0474 & 0.0225 \\
		\hline	NM66-A	&	0.260	&	0.279	&	-13	&	-1	& 0.0393 & 0.0096\\
		\hline		&		&		&		&		& &\\
		\hline	NM44-R	&	0.395	&	0.249	&	33	&	-11	& 0.0976 & 0.0712\\
		\hline	NM45-R	&	0.395	&	0.250	&	33	&	-11	& 0.0975 & 0.0695\\
		\hline	NM46-R	&	0.395	&	0.251	&	33	&	-10	& 0.0977 & 0.0693\\
		\hline	NM54-R	&	0.279	&	0.291	&	-6	&	4	& 0.0200 & 0.0212\\
		\hline	NM55-R	&	0.275	&	0.287	&	-8	&	2	& 0.0237 & 0.0211\\
		\hline	NM56-R	&	0.280	&	0.295	&	-6	&	5	& 0.0195 & 0.0397\\
		\hline	NM64-R	&	0.258	&	0.285	&	-13	&	2	& 0.0405 & 0.0371\\
		\hline	NM65-R	&	0.255	&	0.255	&	-14	&	-9	& 0.0436 & 0.0710\\
		\hline	NM66-R	&	0.258	&	0.282	&	-13	&	1	& 0.0452 & 0.0519\\
		\hline 
		\hline  \textbf{Experiment} & \textbf{0.298} & \textbf{0.280} & && & \\
		\hline
		\end{tabular} 
	\end{center}
\end{table}

\begin{figure}[bt]
\begin{subfigure}[b]{1\textwidth}
\begin{center}
\includegraphics[width=0.75\textwidth]{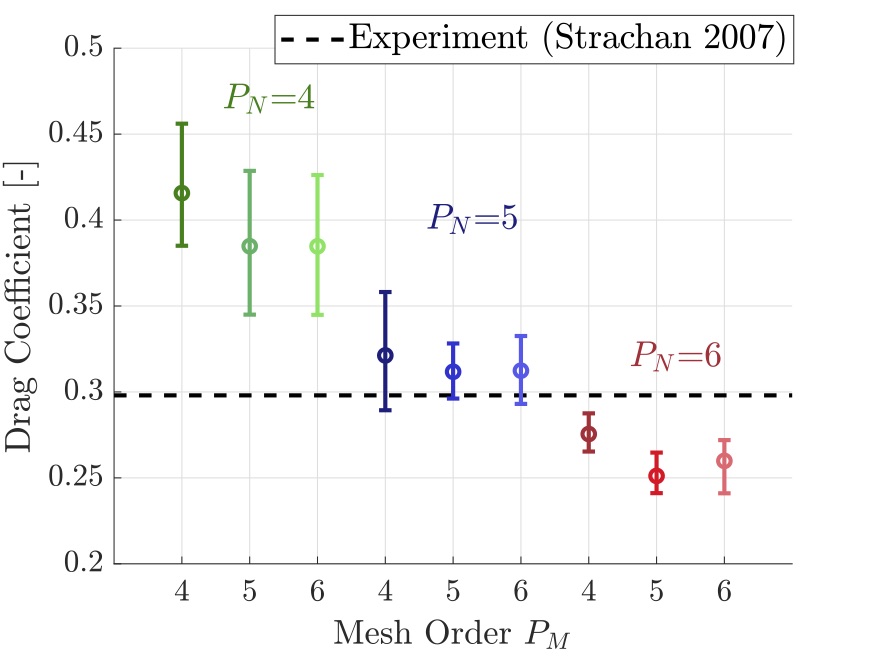}
\caption{Original Mesh}
\label{fig: dragcoef_o}
\end{center}
\end{subfigure}
\begin{subfigure}[b]{1\textwidth}
\begin{center}
\includegraphics[width=0.75\textwidth]{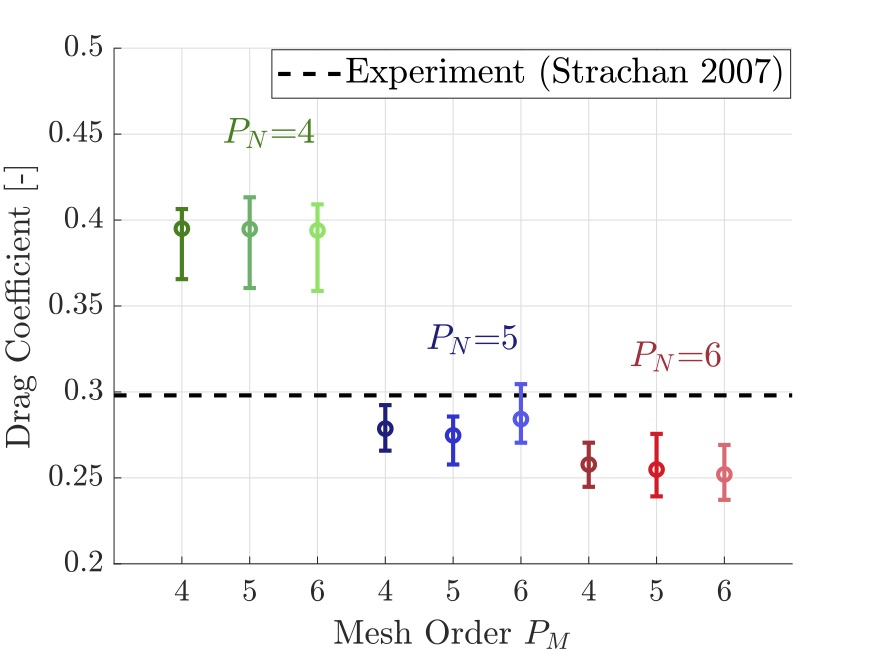}
\caption{Refined Mesh}
\label{fig: dragcoef_r}
\end{center}
\end{subfigure}
\caption{Drag coefficient comparison for \emph{Original} and \emph{Refined} mesh cases, considering proposed high-order meshes and polynomial basis.}
\label{fig: dragcoef}
\end{figure}

\begin{figure}[bt]
\begin{subfigure}[b]{1\textwidth}
\begin{center}
\includegraphics[width=0.75\textwidth]{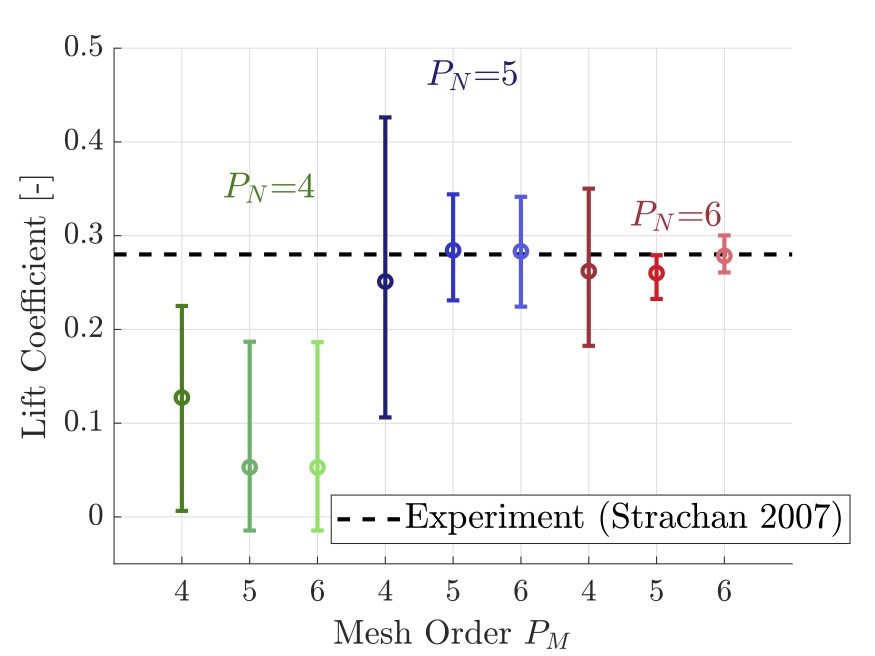}
\caption{Original Mesh}
\label{fig: liftcoef_o}
\end{center}
\end{subfigure}
\begin{subfigure}[b]{1\textwidth}
\begin{center}
\includegraphics[width=0.75\textwidth]{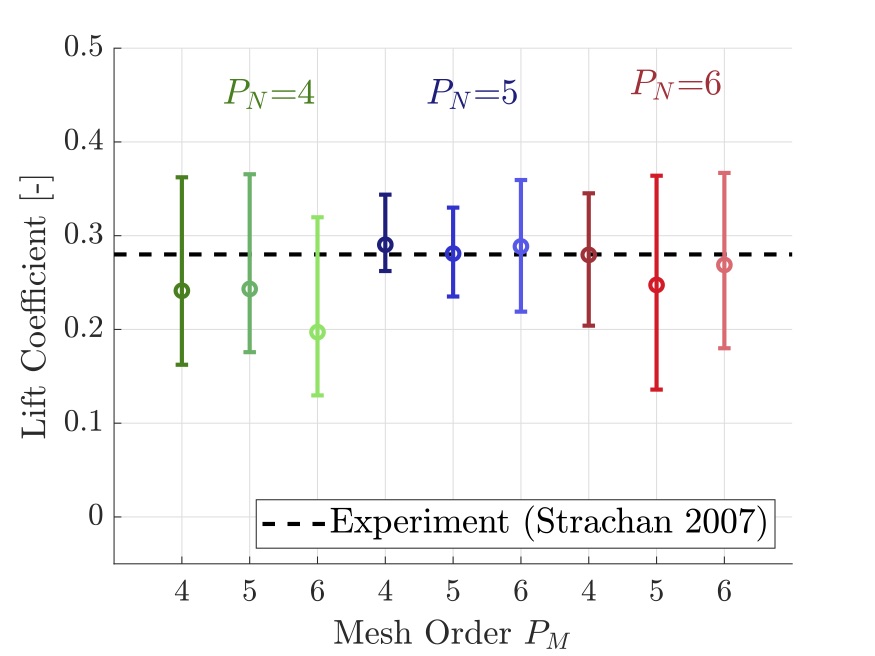}
\caption{Refined Mesh}
\label{fig: liftcoef_r}
\end{center}
\end{subfigure}
\caption{Lift coefficient comparison for \emph{Original} and \emph{Refined} mesh cases, considering proposed high-order meshes and polynomial basis.}
\label{fig: liftcoef}
\end{figure}

In order to better visualize the results of both drag and lift coefficient from the \emph{Original} and \emph{Refined} mesh cases, we present the averaged quantities (dotted line) with minimum and maximum deviation comparing with experimental results (dashed line) from\cite{strachan2007vortex}. The drag coefficient summary is presented in Figure~\ref{fig: dragcoef} and the lift coefficient summary in Figure~\ref{fig: liftcoef}.

The results for the aerodynamic quantities indicate that for a simplified geometry such as the Ahmed body, once passing a threshold value, further increasing the mesh order has small influence on the results. For this type of bluff bodies, most of the relevant flow structures are generated from sharp edges of the geometry, such as the slant. The only curved surface is the frontal stagnation, where we see enhancements on capturing the curvature as the $P_M$ increases. We select to have the highest high-order mesh resolution ($P_M$ = \nth{6} order) as our reference for curvature mesh for flow structure results to be presented and further application on the diffuser test cases, as also employed by \cite{buscariolo2019spectralhp} and similar automotive studies. This study also highlights that for the \emph{Original} mesh case, we observe similar values for drag and lift coefficients for $P_M = 5$ and $P_M = 6$. It indicates that further increase on $P_M$ will not affect results. On the \emph{Refined} mesh case, drag coefficient values are similar for every $P_M$, however the lift coefficient improves as $P_N$ for the solution increases.

When analysing numerical results, the outcome for the best numerical methodology are between \emph{Original} and \emph{Refined} meshes with $P_N=5$ and $P_N=6$ resolution, respectively for drag and lift coefficients. A point to consider is that for $P_N=5$, we noticed that increasing the h-refinement from \emph{Original} to \emph{Refined}, the result changed from over-predicted to under-predicted. Results for drag considering $P_N=6$ resolution maintained similar value for both meshes, indicating consistency. 

\begin{figure}[hbt]
	\begin{center}
	\includegraphics[width=1.0\textwidth]{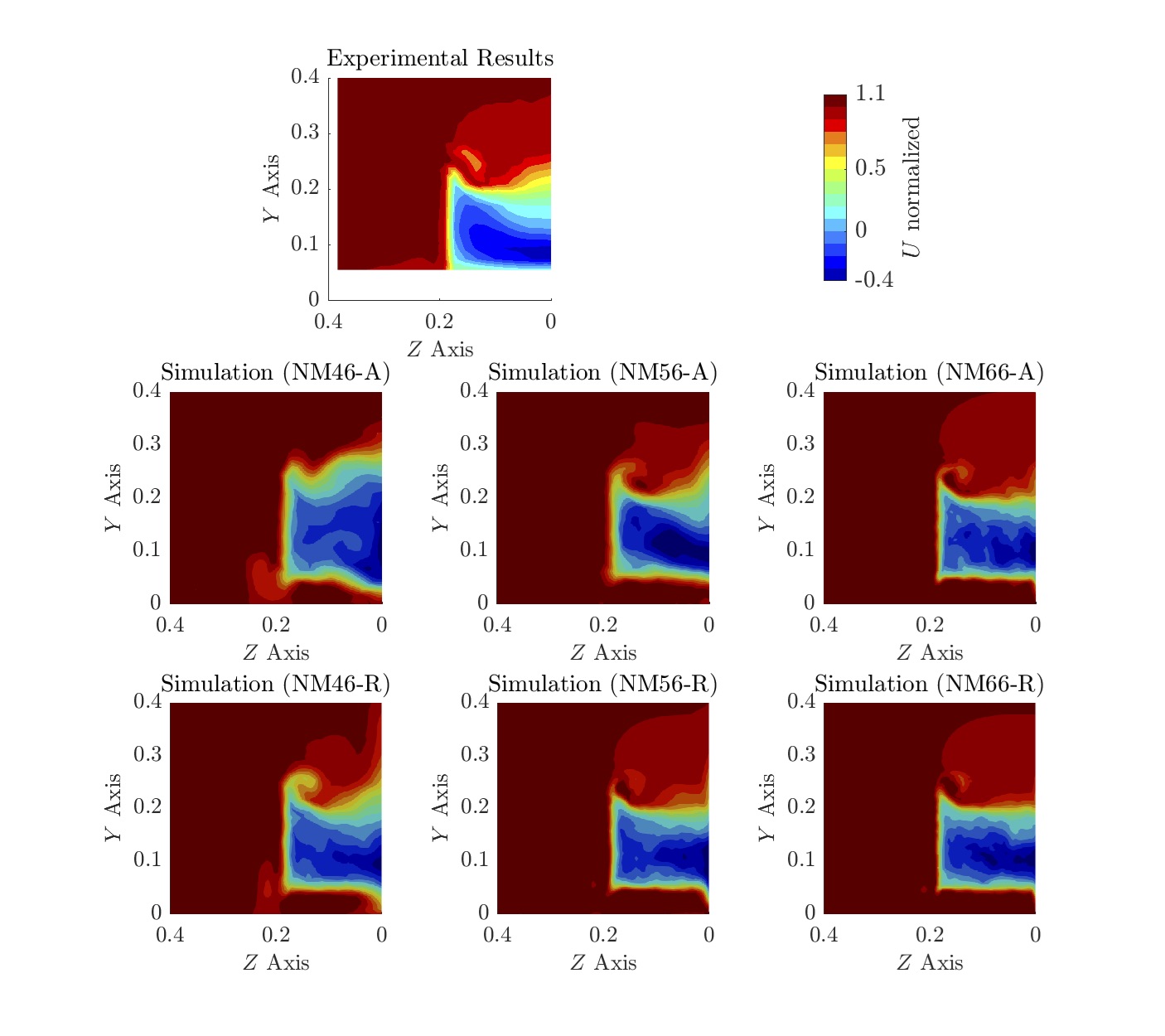}
	\caption{Comparison of normalized streamwise velocity $U$ between experiments of \cite{lienhart2003flow} (top) and computational simulations on plane $ZY$ at $X/L = 0.077$.}
	\label{fig:uvel_ahmed}
	\end{center}
\end{figure}

In terms of flow structure comparison, we point out that the main reference used for the aerodynamic quantities (\cite{strachan2007vortex}) has an upper support over the body to allow the moving ground condition, which the upper support contributions for the drag forces being empirically deducted from the obtained total value. However, the influences on flow features will still remain, possibly leading to weaker vortices over the slant when comparing to results without the support. Following the findings of \cite{strachan2007vortex} that the upper strut can interfere with the vortices’ intensity, we are comparing aerodynamics flow visualization with results from \cite{lienhart2003flow} (without strut), similar as this work. We present a comparison of the normalized $U$ on the plane ZY at $X/L = 0.077$ in Figure~\ref{fig:uvel_ahmed}, aiming to compare wake structures in the flow. 

Analysing the flow, \emph{Refined} mesh with $P_N=5$ and $P_N=6$ cases qualitatively closely correlate to experimental results, with better definition of flow structures and larger contour spectrum. Especially for $P_N=6$ resolution, capturing similar shape of the experimental reference.

The main conclusion here presented is that the minimum resolution for this mesh setup to reproduce similar features and aerodynamic quantities is the \emph{Refined} configuration combined with \nth{6} high-order mesh ($P_M$ = 6) and polynomial expansion of \nth{6} order accuracy ($P_N=6$). The \emph{Refined} configuration combined with polynomial expansion of \nth{6} order accuracy leads to $y^{+}\leq1$ over the body walls, giving support to previous statement and is applied to the all further simulations in this study.

\begin{figure}[hbt]
	\begin{center}
	\includegraphics[width=0.7\textwidth]{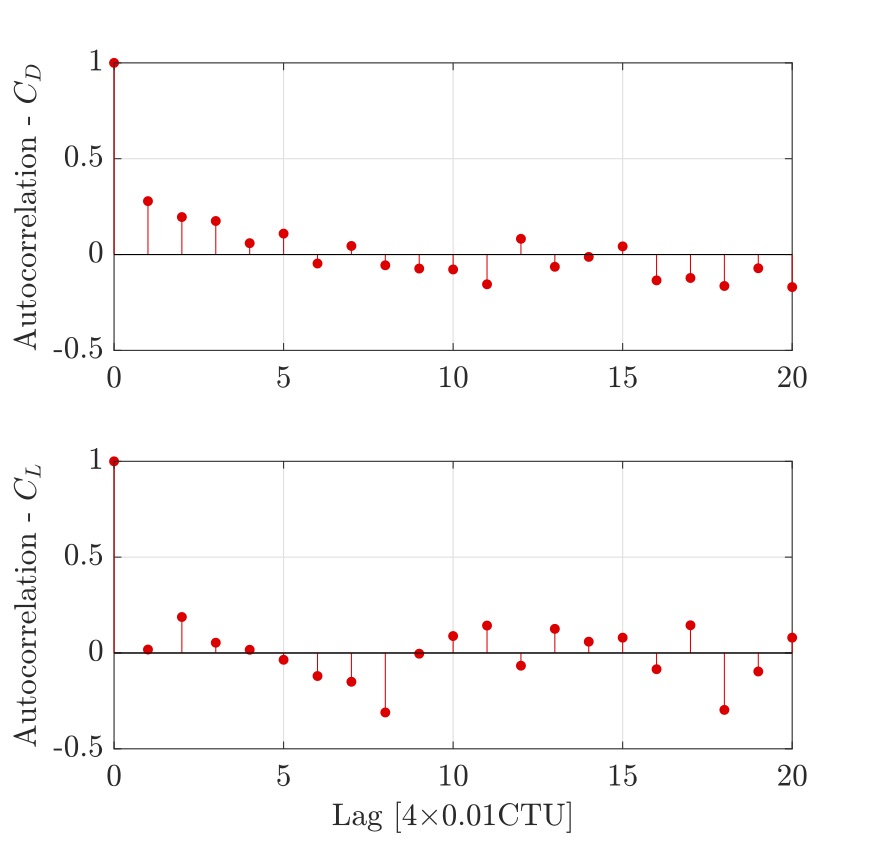}
	\caption{Autocorrelation for the re-sampled unsteady drag and lift coefficient}
	\label{fig: Autocorrelation}
	\end{center}
\end{figure}

In order to justify the reliability of the aerodynamic quantities results of the selected resolution, we applied the method proposed by \cite{islam2017detailed} and further applied by \cite{luckhurst2019computational}. Drag and lift coefficient data was resampled to remove the statistical dependence in the unsteady signal, between the 5 to 7 CTU, shown in Figure\ref{fig: Autocorrelation}.

\begin{figure}[hbt]
	\begin{center}
	\includegraphics[width=0.7\textwidth]{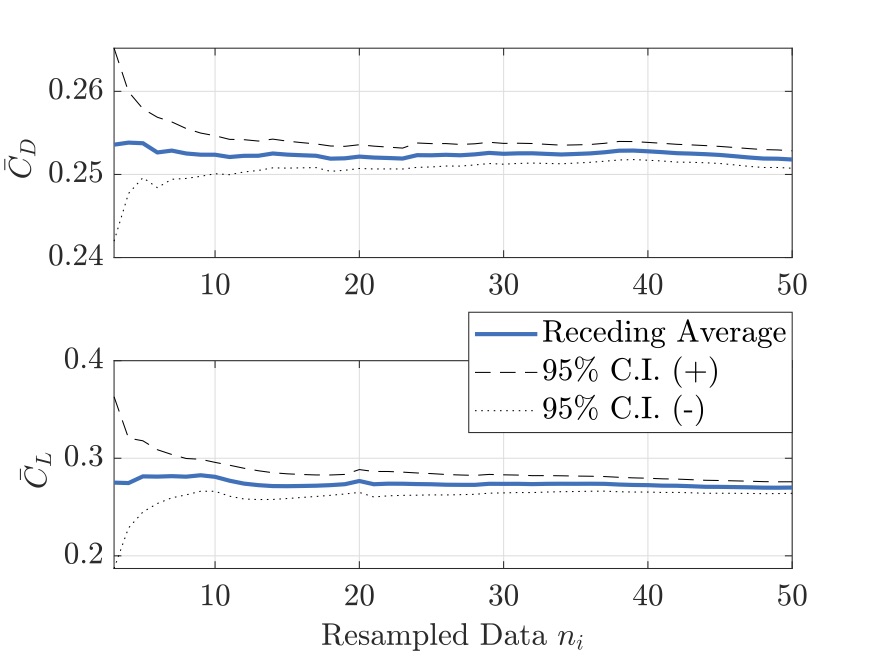}
	\caption{Receding average with 95\% confidence intervals}
	\label{fig: RecedingAverageCI}
	\end{center}
\end{figure}

Considering the resampled data for both drag and lift coefficients, the receding average of was generated with a 95\% confidence interval and presented in Figure\ref{fig: RecedingAverageCI}. It indicates that significance of the autocorrelation is removed and the data from 5 to 7 CTUs is statiscally independent.

The results here presented are also in accordance with current literature on the use of LES solutions on the Ahmed body in terms of the drag coefficient, ranging from $6\%$ to $16\%$, according to \cite{serre2013simulating} studies, but showing great agreement in terms of lift coefficient. Uncertainties of the experiment might also explain the gap between computational results, such as the use of the upper strut. The strut influence on the drag coefficient results of \cite{strachan2007vortex} is approximately 15\% when compared to the study of \cite{graysmith1994comparisons}. This study also highlights improvements in terms of also capturing correct flow structures and intensities. From previous reference of spectral methods applied to this bluff body, there is a reducing from from 44\% (\cite{serre2013simulating} to 13\% on the drag coefficient value.



\section{Simulation of the Ahmed body with diffuser}
\label{sec:simulation}

\begin{figure}[hbt]
	\begin{center}
	\includegraphics[width=1\textwidth]{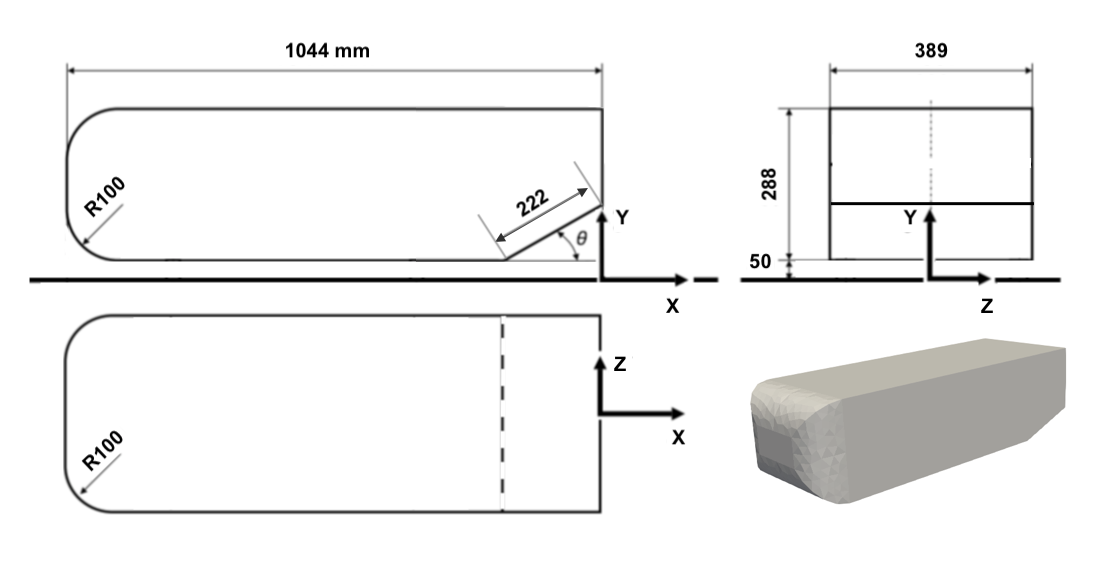}
	\caption{Schematic drawing of the Ahmed body squared-back equipped with rear underbody diffuser.}
	\label{fig:draw_ahmed0_dif}
	\end{center}
\end{figure}

\begin{figure}[hbt]
	\begin{center}
	\includegraphics[width=1\textwidth]{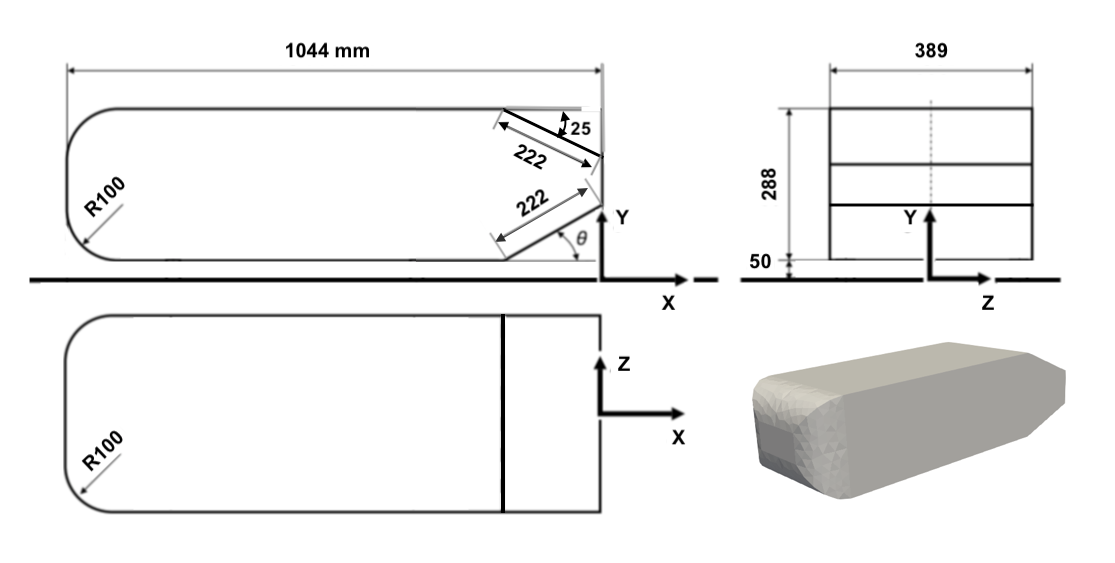}
	\caption{Schematic drawing of the Ahmed body with slant angle of 25$^{\circ}$ equipped with rear underbody diffuser.}
	\label{fig:draw_ahmed25_dif}
	\end{center}
\end{figure}

With the numerical methodology validated on the classical Ahmed body, the following step is the application on a new variant of the Ahmed body equipped with underbody diffuser. Diffuser length $D_L$ is set to be at a fixed value, which is the same as the upper slant length $S_L$ of 222 mm, regardless of the inclination angle changes. The influence of the diffuser is evaluated in two variants of the classical Ahmed body: 0$^\circ$ slant (or squared-back) and 25$^\circ$ slant angle, as illustrated in Figures \ref{fig:draw_ahmed0_dif} and \ref{fig:draw_ahmed25_dif}. The diffuser angle evaluated for was changed from 0$^\circ$ to 50$^\circ$ in increments of 10$^\circ$ with an additional case considering the angle of 5$^\circ$.

\begin{figure}[hbt]
	\begin{center}
	\includegraphics[width=0.8\textwidth]{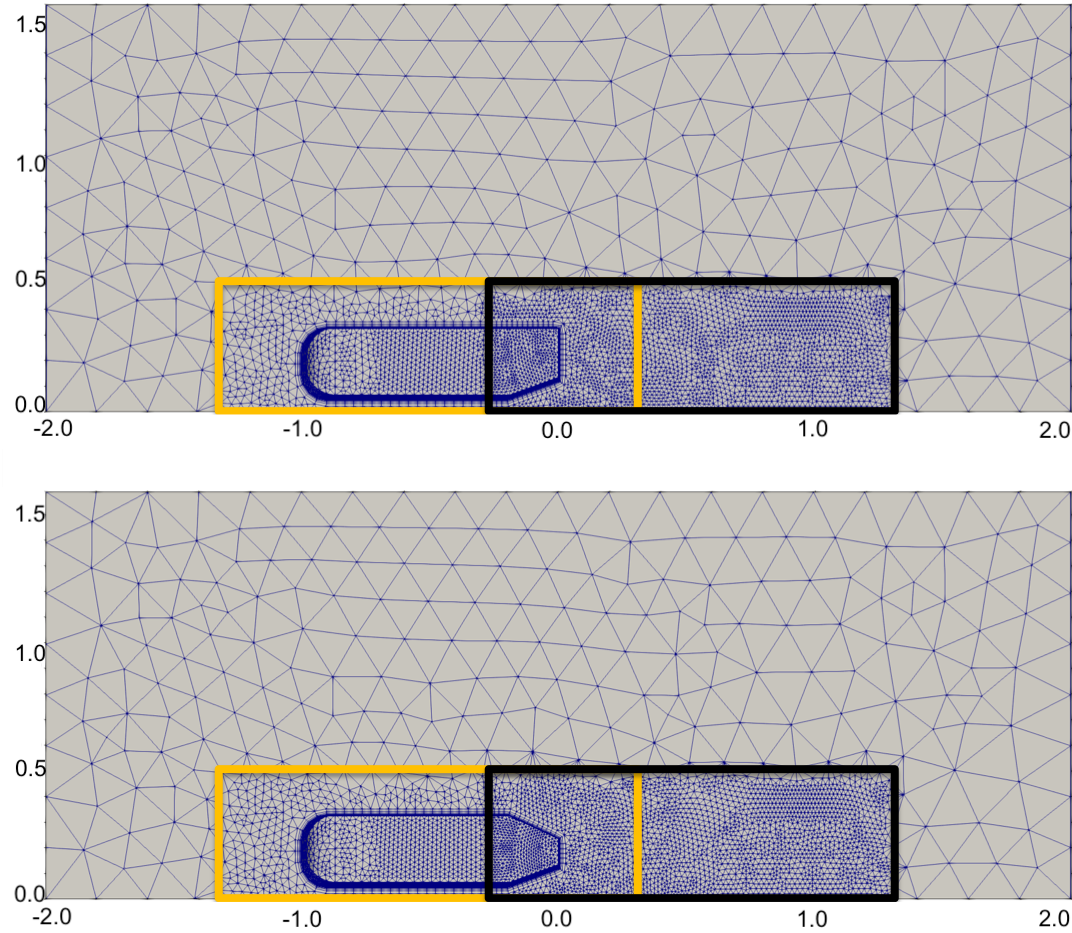}
	\caption{Mesh refinement regions on both Ahmed bodies squared back (up) and with slant angle of 25$^{\circ}$. Refinement region highlighted in yellow is defined as the Ahmed Body refinement and region highlighted in black is defined as Wake Refinement.}
	\label{fig:mesh_ref}
	\end{center}
\end{figure}

All simulation cases were performed based on the methodology established on previous chapter and in the study of \cite{filipe2018ahmed}, considering \emph{Refined} mesh parameters, with \nth{6} high-order mesh ($P_M$ = 6) and polynomial expansion of \nth{6} order accuracy ($P_N=6$), corresponding to the NM66-R case on the validation study. The volumetric mesh also incorporates the two refinement zones (Ahmed body and wake refined) previously and illustrated for each body style in Figure~\ref{fig:mesh_ref}.

With this simulation setup using a polynomial expansion for the solution $P_N = 6$, we are able to increases the resolution of the solution by converting a relatively coarse mesh with 250,000 (h-type equivalent) elements into 19.8 million degrees of freedom (DOF) per variable. Boundary conditions and Reynolds number for the simulation cases with diffuser are the same applied for the numerical methodology validation study and following the work of \cite{strachan2007vortex}.





\subsection{Results}
\label{subsec:5}

The key findings of the study are presented as follows. We initially present a comparison of drag and lift coefficients for both Ahmed body styles when the diffuser angles changes. Both quantities are averaged from the \nth{5} to the \nth{7} convective length where we assure to have a fully converged physical solution. Flow structures comparison for the planes $X/L =0$ and $X/L =0.096$ and wall shear stress lines are both averaged from the same period and presented in order to complement the findings.

\subsubsection{Drag coefficient results}
\label{subsubsec:drag}

\begin{figure}[hbt]
	\begin{center}
	\includegraphics[width=1\textwidth]{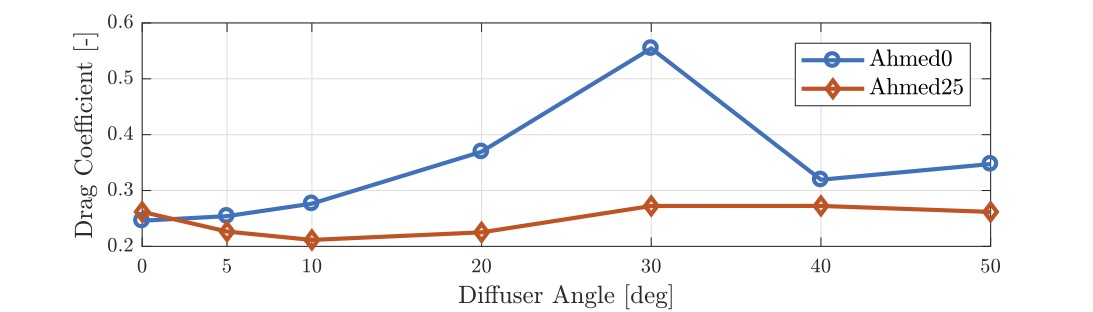}
	\caption{Drag coefficient comparison for Ahmed Body squared-back (blue line) and 25$^{\circ}$ slant inclination (orange line) considering standard configuration and evaluated diffuser angles: 5$^{\circ}$, 10$^{\circ}$, 20$^{\circ}$, 30$^{\circ}$, 40$^{\circ}$ and 50$^{\circ}$.}
	\label{fig:avgdrag_f}
	\end{center}
\end{figure}

Drag coefficient results with both Ahmed body slant angles are presented in Figure~\ref{fig:avgdrag_f}. Ahmed body squared back results indicate that the drag coefficient initially rises as the diffuser angle increases, reaching the maximum value at the diffuser angle of 30$^{\circ}$. The rising drag trend suddenly breaks for diffuser angles higher than 30$^{\circ}$, where results for further angles are similar to the diffuser with 20$^{\circ}$ inclination. Such behaviour indicates similar trends as earlier verified in studies on Ahmed body slant angle variations of \cite{ahmed1984some}, \cite{lienhart2003flow} and \cite{strachan2007vortex}. The drag breaking point for the diffuser angle is similar to slant angle which is an indication of flow structure change on the diffuser region and is further discussed below. We conclude that any diffuser added to the Ahmed body squared back has a penalty in terms of drag performance. 

The drag coefficient results for Ahmed body with 25$^{\circ}$ slant angle show different trend from the squared back case. Applying the diffuser with angles of 5$^{\circ}$, 10$^{\circ}$ and 20$^{\circ}$ angle leads to drag reduction with the optimum angle at 10$^{\circ}$. For the other diffuser angles, the drag coefficient recovers to a similar value of the Ahmed body with 25$^{\circ}$ slant angle without a diffuser. We also notice that the drag performance enhancement region might be related to flow behaviour change on the diffuser region, which is further presented. Diffuser application has no negative impact on drag coefficient for this Ahmed body style.

\subsubsection{Lift coefficient results}
\label{subsubsec:lift}

\begin{figure}[hbt]
	\begin{center}
	\includegraphics[width=1\textwidth]{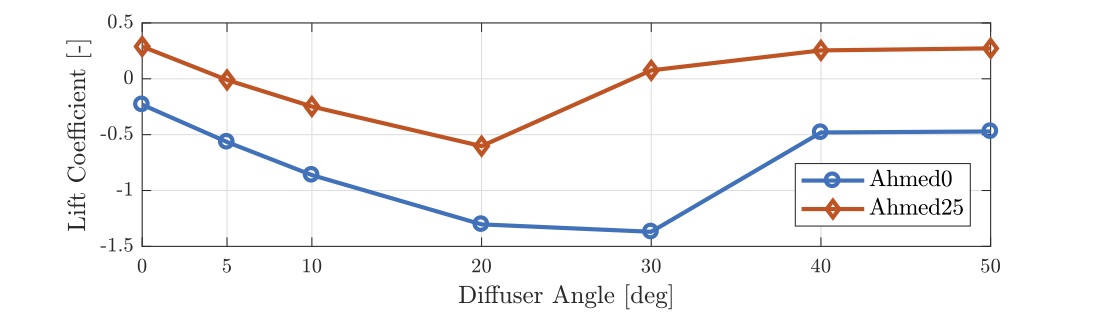}
	\caption{Lift coefficient comparison for Ahmed Body squared-back (blue line) and 25$^{\circ}$ slant inclination (orange line) considering standard configuration and evaluated diffuser angles: 5$^{\circ}$, 10$^{\circ}$, 20$^{\circ}$, 30$^{\circ}$, 40$^{\circ}$ and 50$^{\circ}$.}
	\label{fig:avglift_f}
	\end{center}
\end{figure}

Lift coefficient values for both Ahmed body cases at different diffuser angles are presented in Figure~\ref{fig:avglift_f}. When analysing the lift coefficient results for the Ahmed body squared back, downforce enhancement is obtained as the diffuser angle increases from 0$^{\circ}$, reaching maximum downforce value at 30$^{\circ}$. The downforce trend breaking phenomenon is observed for diffuser angles higher than 30$^{\circ}$, similar as observed for the drag coefficient. For diffuser angles higher than 30$^{\circ}$, downforce force values are similar to the diffuser of 5$^{\circ}$, indicating saturation of downforce enhancement with this diffuser geometry.

Ahmed body with 25$^{\circ}$ slant inclination lift coefficient results shows similar trend of previous squared-back results. We've noticed the positive lift on the baseline case without diffuser, which might compromise gripping on performance and race cars. By implementing the diffuser, downforce performance starts to increase, where the first proposed diffuser angle of 5$^{\circ}$ take the lift coefficient to equilibrium. Downforce increment is noticed until the diffuser angle of 20$^{\circ}$, where the maximum performance is reached. The diffuser loses its performance for diffuser angle of 30$^{\circ}$ and above, where the first has performance similar to the 5$^{\circ}$ whereas the 40$^{\circ}$ and 50$^{\circ}$ diffusers have similar performance as the baseline.

\subsubsection{Flow Features Analysis}
\label{subsubsec:flow}

Comparative results for the flow structures found on the Ahmed body equipped with a diffuser are now presented and discussed. The comparisons are presented at two planes, one at $X/L =0$, where the back end of the Ahmed body is placed and one downstream at $X/L =0.096$, in order to evaluate how the flow structures develop as they separate from the body. Contours of Q-criterion, $U$ (streamwise) and $V$ (vertical) velocities are provided for the inspection planes, aiming to identify flow structures and define its interactions with the rear wake. We also provide a flow topology comparison on the diffuser surface by plotting averaged wall shear stress lines from the \nth{5} to the \nth{7} convective length, same period as the flow quantities and aerodynamic forces. Results for the Ahmed body squared-back are firstly presented, followed by the Ahmed body with 25$^{\circ}$ slant angle with all plane views observed from downstream. We define the simulation cases by the abbreviation of diffuser angle, DA, followed by the inclination angle so diffuser angle of 30$^{\circ}$ is defined as DA30.

\subsubsection{Ahmed body squared back with diffuser}
\label{subsubsec:ahmed0flow}

\begin{figure}[bt]
\begin{subfigure}[b]{1\textwidth}
\begin{center}
\makebox[\textwidth][c]{\includegraphics[width=1.3\textwidth]{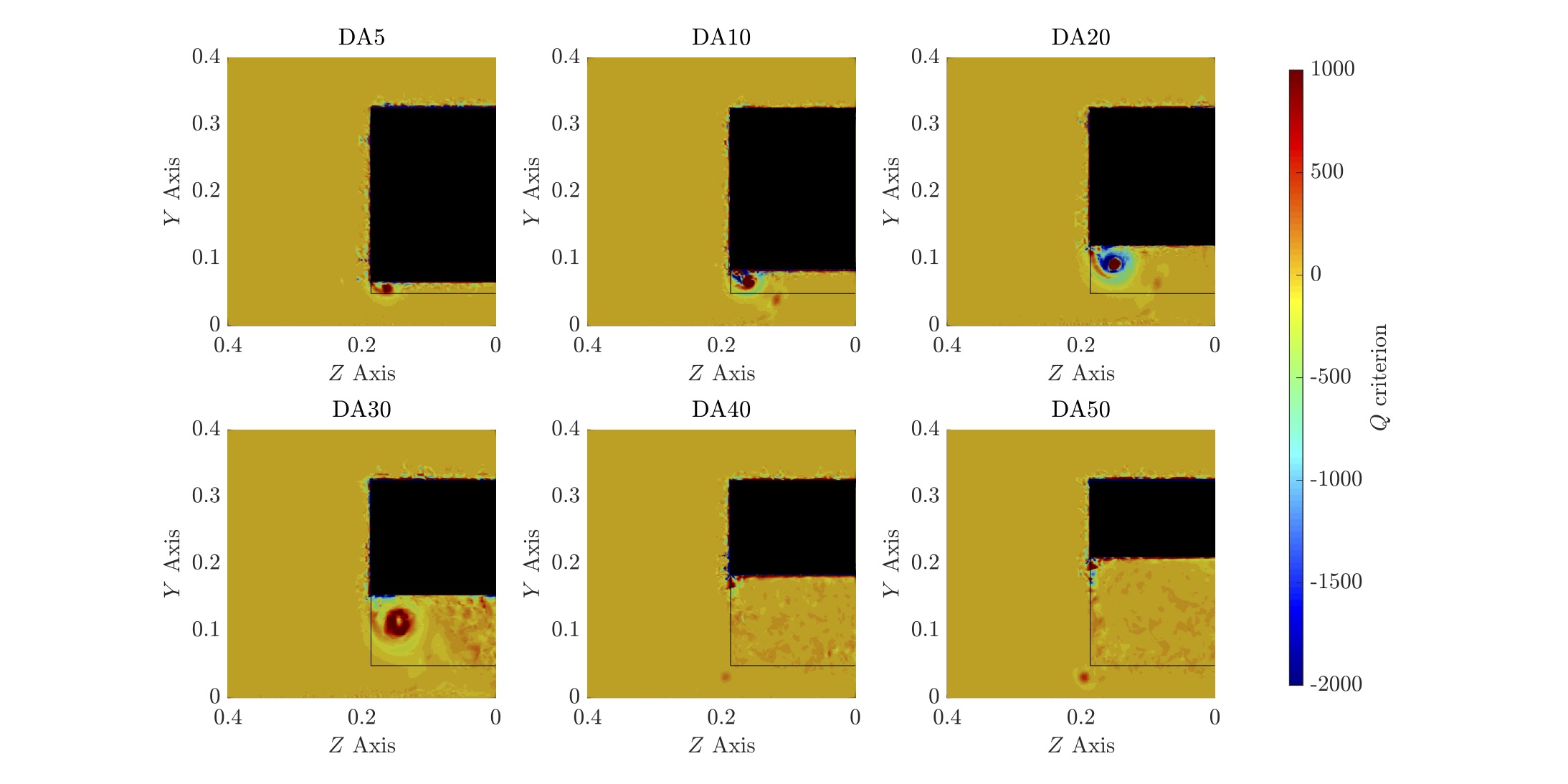}}%
\caption{Plane $X/L =0$}
\label{fig:Ahmed0_Plane0_Q}
\end{center}
\end{subfigure}
\begin{subfigure}[b]{1\textwidth}
\begin{center}
\makebox[\textwidth][c]{\includegraphics[width=1.3\textwidth]{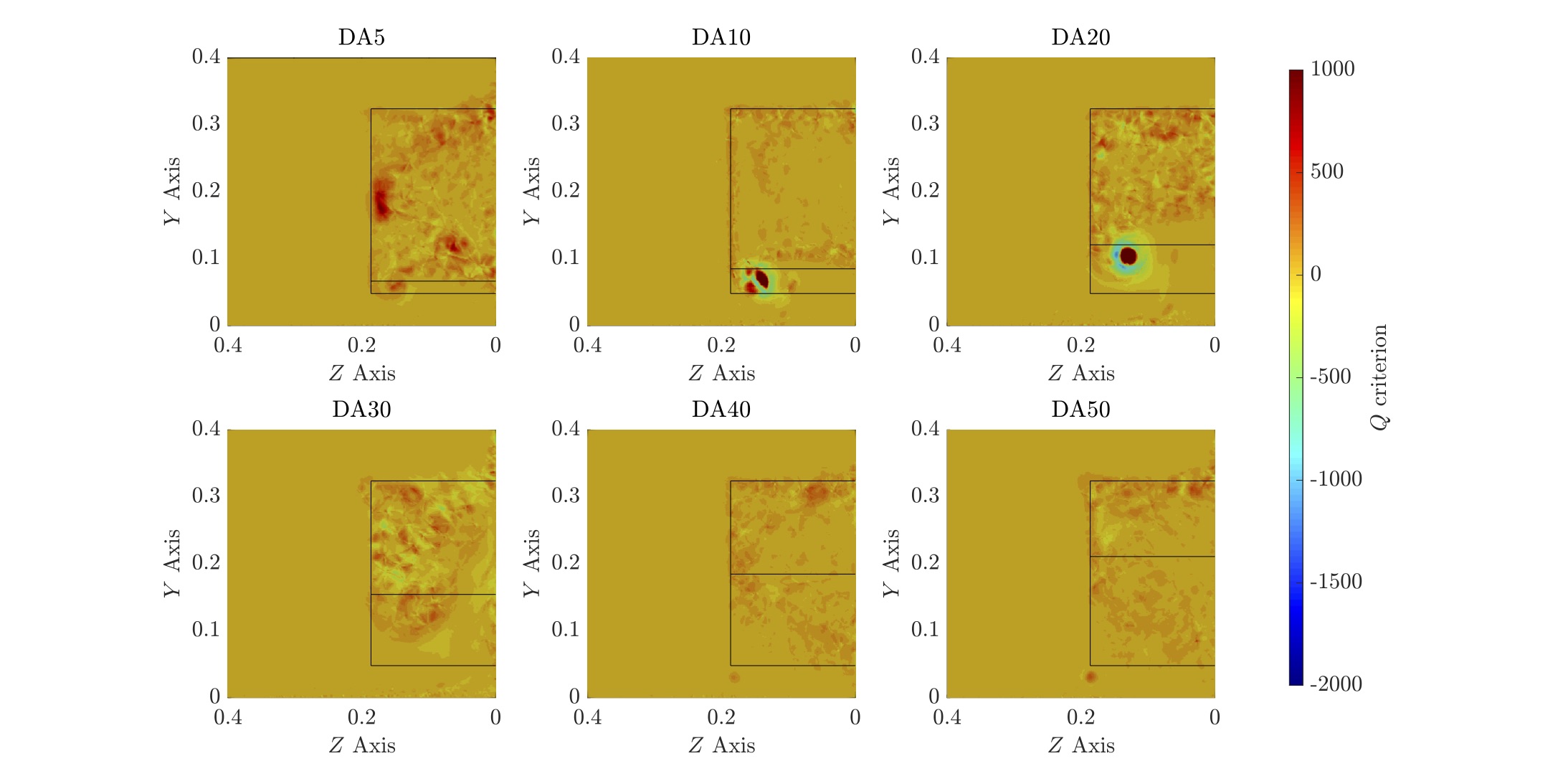}}%
\caption{Plane $X/L =0.096$}
\label{fig:Ahmed0_Plane096_Q}
\end{center}
\end{subfigure}
\caption{Contours of Q-Criterion for the Ahmed body squared-back considering diffuser angle of 5$^\circ$ (DA5), 10$^\circ$ (DA10), 20$^\circ$ (DA20), 30$^\circ$ (DA30), 40$^\circ$ (DA40) and 50$^\circ$ (DA50) for planes $X/L =0$ and $X/L =0.096$.}
\label{fig:Ahmed0_Q}
\end{figure}

The first consideration of Figure~\ref{fig:Ahmed0_Plane0_Q} shows a vortex arising from side of the diffuser up to the angle of 30$^{\circ}$. The vortex intensity and core size increase as the diffuser angle becomes more inclined. A wake structure can also be noticed for the DA30 case, indicating that there is separated flow on the diffuser. Similar flow behaviour is observed on the classical Ahmed body slant variation experiment for slant angles ranging from 12.5$^{\circ}$ to 30$^{\circ}$. For diffuser angles above 30$^\circ$ only a few structures are observed, such as weak vortices on the lower part part of the body, near the floor. This is an indicative of fully separated flow, however the Q-Criterion image alone is not conclusive.

Moving to plane $X/L = 0.096$, shown Figure~\ref{fig:Ahmed0_Plane096_Q}, the main difference noticed is on DA30 case where the diffuser vortex has merged with the rear wake. We observed that the diffuser vortex in DA5, DA10 and DA20 cases are weaker and shifting both upwards and in the spanwise direction towards the centre plane.

\begin{figure}[bt]
\begin{subfigure}[b]{1\textwidth}
\begin{center}
\makebox[\textwidth][c]{\includegraphics[width=1.3\textwidth]{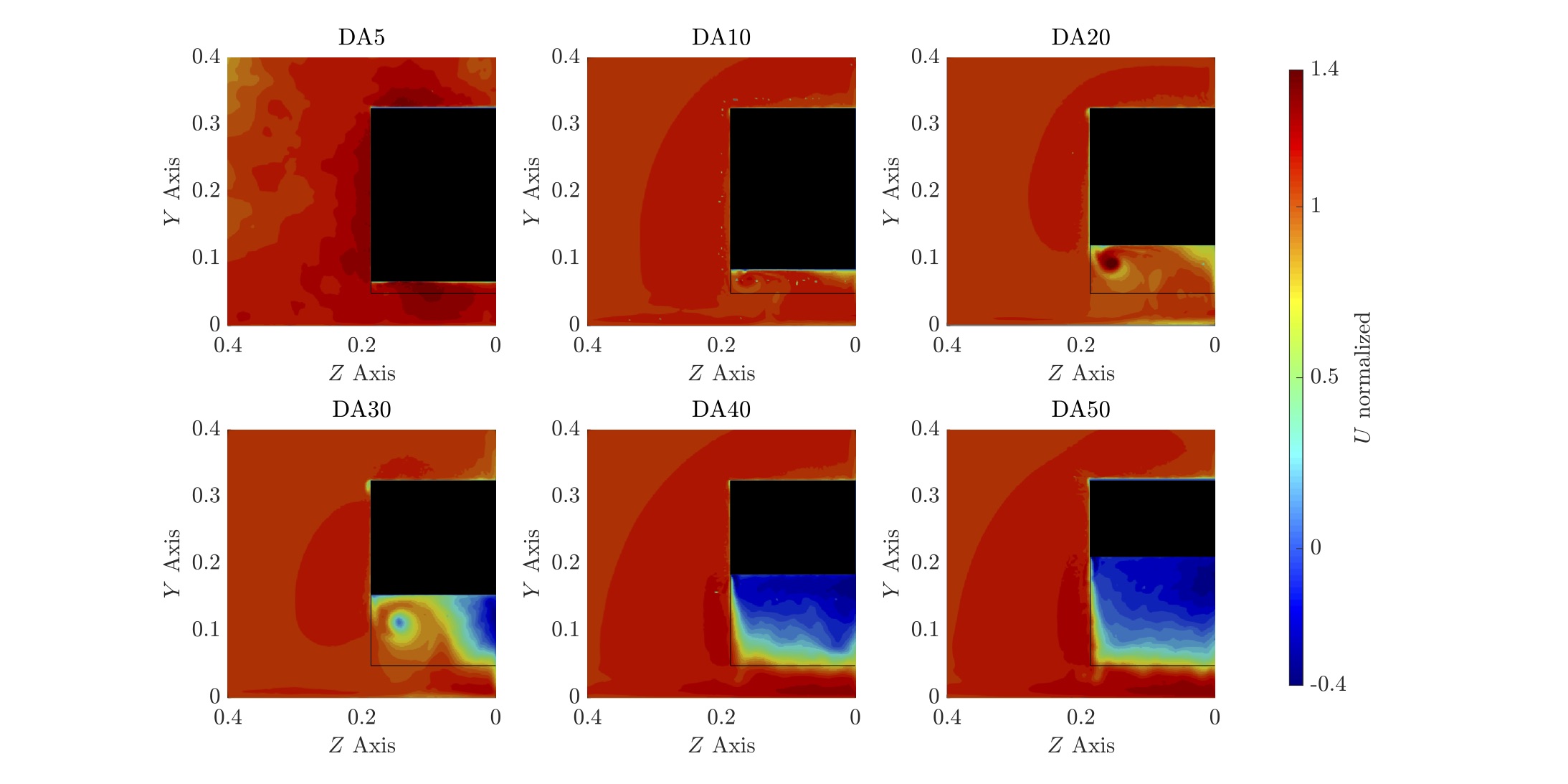}}%
\caption{Plane $X/L =0$}
\label{fig:Ahmed0_Plane0_U}
\end{center}
\end{subfigure}
\begin{subfigure}[b]{1\textwidth}
\begin{center}
\makebox[\textwidth][c]{\includegraphics[width=1.3\textwidth]{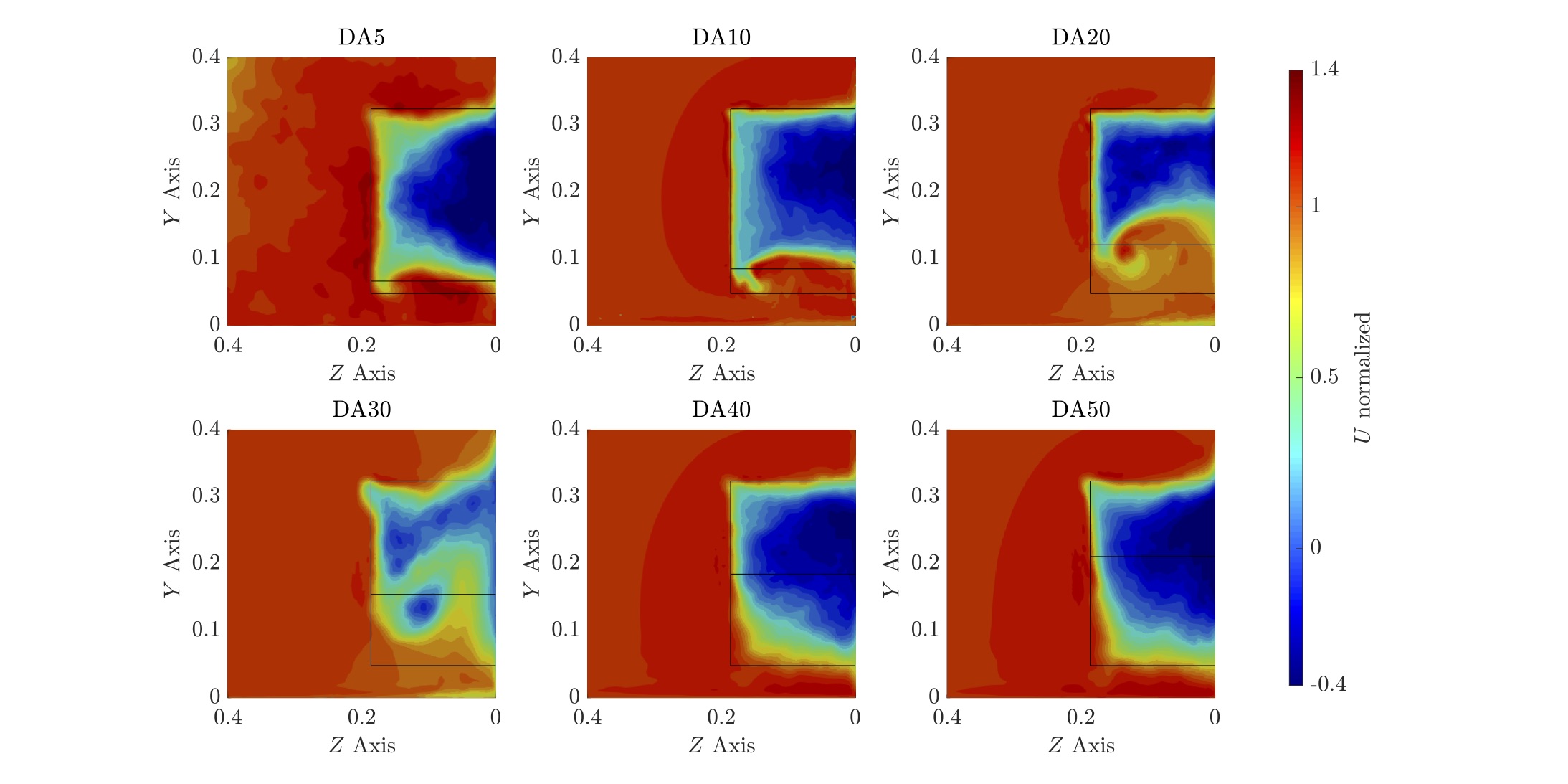}}%
\caption{Plane $X/L =0.096$}
\label{fig:Ahmed0_Plane096_U}
\end{center}
\end{subfigure}
    \caption{Contours of normalized streamwise velocity $U$ for the Ahmed body squared-back considering diffuser angle of 5$^\circ$ (DA5), 10$^\circ$ (DA10), 20$^\circ$ (DA20), 30$^\circ$ (DA30), 40$^\circ$ (DA40) and 50$^\circ$ (DA50) for planes $X/L =0$ and $X/L =0.096$.}
\label{fig:Ahmed0_U}
\end{figure}

Flow separation is observed for DA40 and DA50 case on plane $X/L = 0$. The large negative velocity area shown in Figure~\ref{fig:Ahmed0_Plane0_U} indicates that the flow is already separated at the outlet of the diffuser. The slight higher drag for DA50 is explained by a more intense contour of negative $U$ velocity when comparing to DA40. The negative velocity area on DA30 at mid-span, shows a combination of flow separation with vortex generation. With both flow features, the DA30 case can characterized as a highly energetic flow. 

Flow structures evolution is presented in Figure~\ref{fig:Ahmed0_Plane096_U}, reaching the plane $X/L = 0.096$. Low velocity zones are observed on the rear vertical portion of the body, with a different contour position for DA30. The DA30 shifts the base wake upward as it moves to a inner spanwise direction. The diffuser vortex intensity is probably the main cause of this translation and $V$ velocity results are next presented to complement the explanation the this phenomenon.

\begin{figure}[bt]
\begin{subfigure}[b]{1\textwidth}
\begin{center}
\makebox[\textwidth][c]{\includegraphics[width=1.3\textwidth]{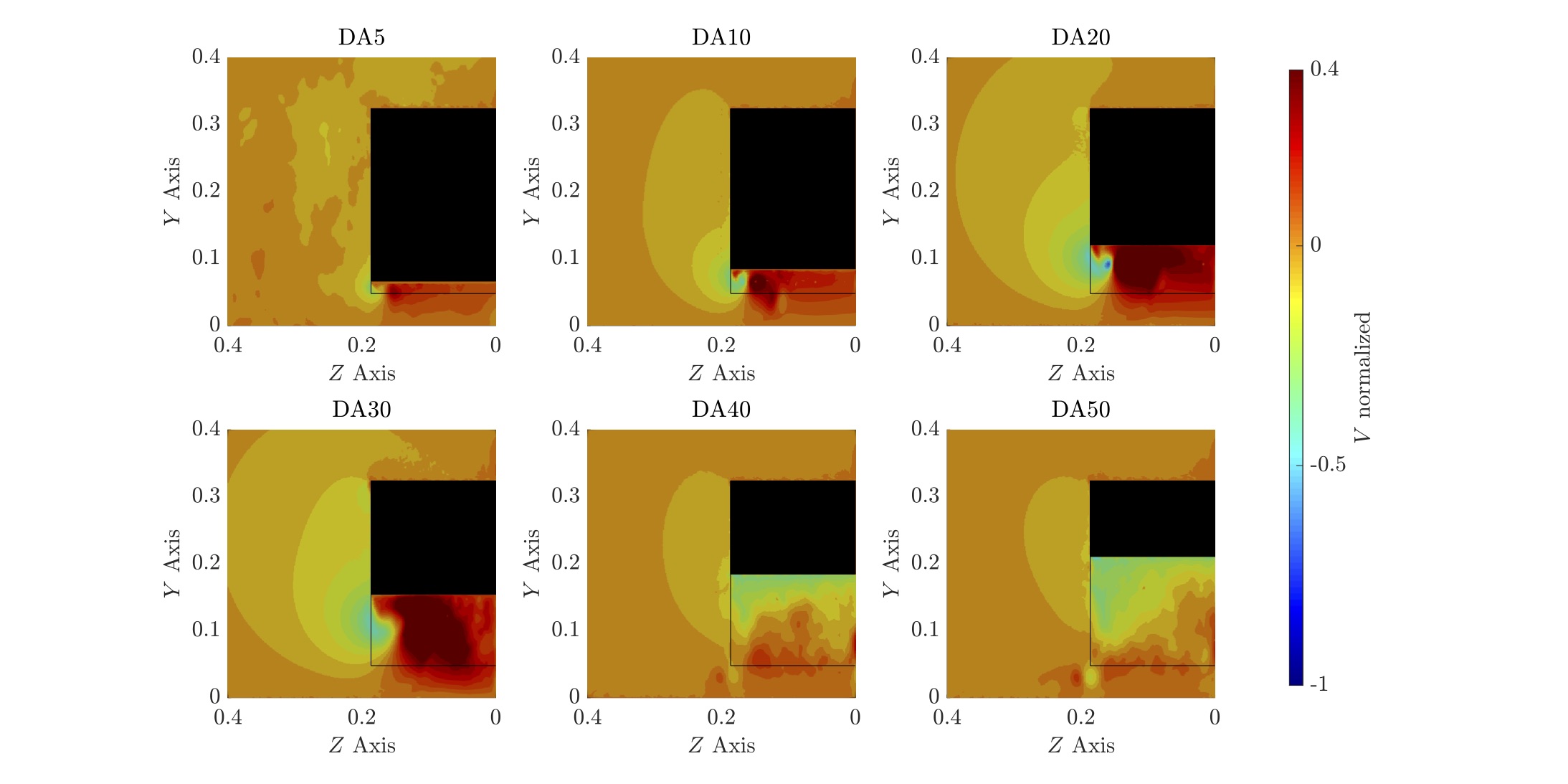}}%
\caption{Plane $X/L =0$}
\label{fig:Ahmed0_Plane0_V}
\end{center}
\end{subfigure}
\begin{subfigure}[b]{1\textwidth}
\begin{center}
\makebox[\textwidth][c]{\includegraphics[width=1.3\textwidth]{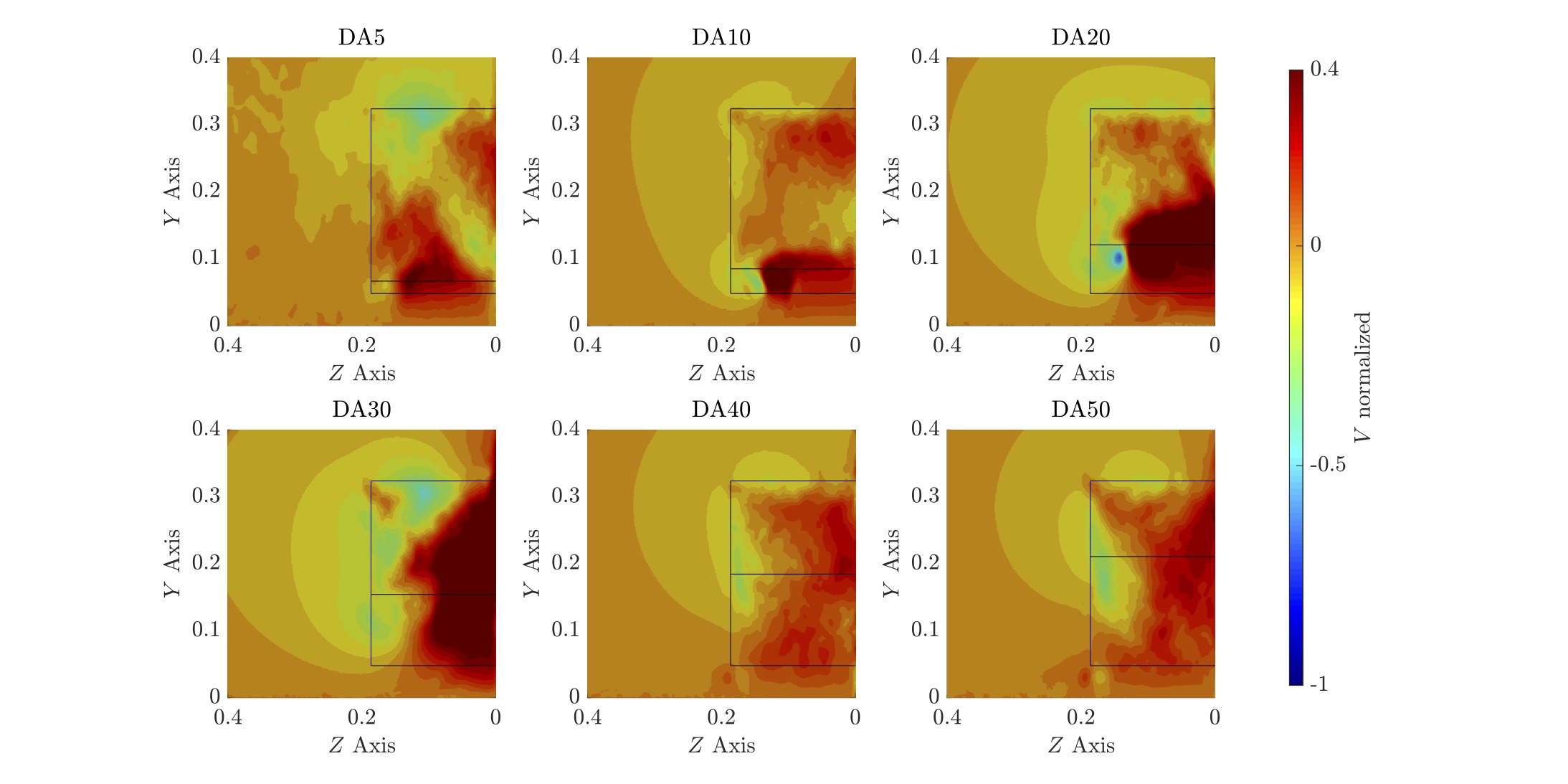}}%
\caption{Plane $X/L =0.096$}
\label{fig:Ahmed0_Plane096_V}
\end{center}
\end{subfigure}
    \caption{Contours of normalized vertical velocity $V$ for the Ahmed body squared-back considering diffuser angle of 5$^\circ$ (DA5), 10$^\circ$ (DA10), 20$^\circ$ (DA20), 30$^\circ$ (DA30), 40$^\circ$ (DA40) and 50$^\circ$ (DA50) for planes $X/L =0$ and $X/L =0.096$.}
\label{fig:Ahmed0_V}
\end{figure}

From normalized vertical velocity $V$ contours in Figure~\ref{fig:Ahmed0_V}, we extract the diffuser vortex rotation direction. The diffuser vortex rotates anti-clockwise, as the most inner spanwise vortex component has positive vertical velocity and the outer negative. Vertical velocity contour for DA30 on plane $X/L = 0.096$ explain the wake moving upwards on the spanwise direction from Figure~\ref{fig:Ahmed0_Plane096_U}. A strong positive vertical velocity zone is observed at the mid-span behind both the diffuser and back of the Ahmed body. 

\begin{figure}[hbt]
\begin{subfigure}[b]{0.33\textwidth}
\begin{center}
\includegraphics[width=1\textwidth]{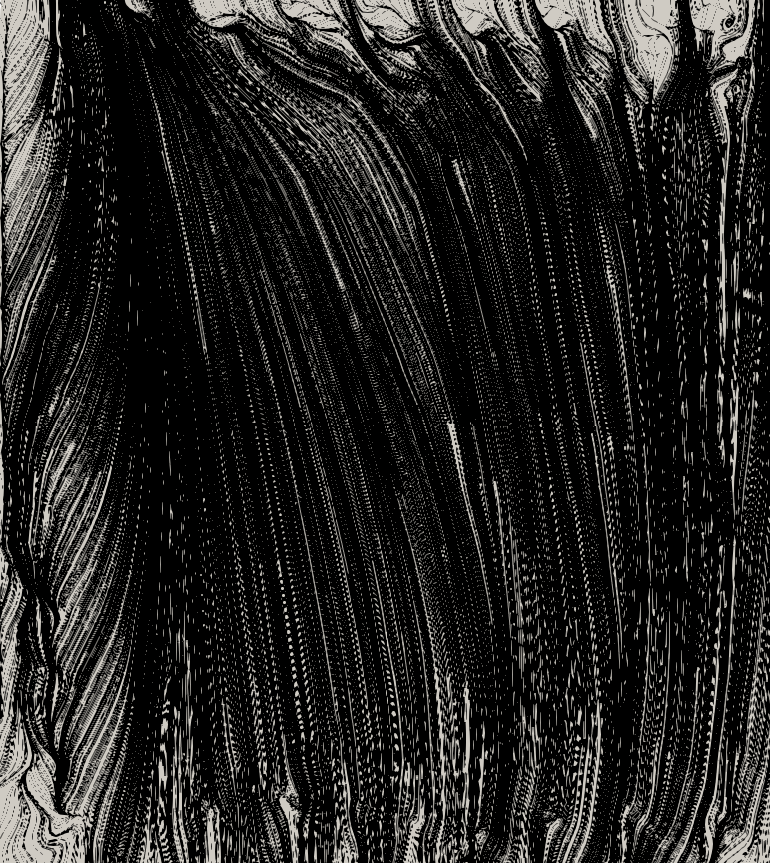}
\caption{DA5}
\label{fig:Ahmed0_wss_DA5}
\end{center}
\end{subfigure}
\begin{subfigure}[b]{0.33\textwidth}
\begin{center}
\includegraphics[width=1\textwidth]{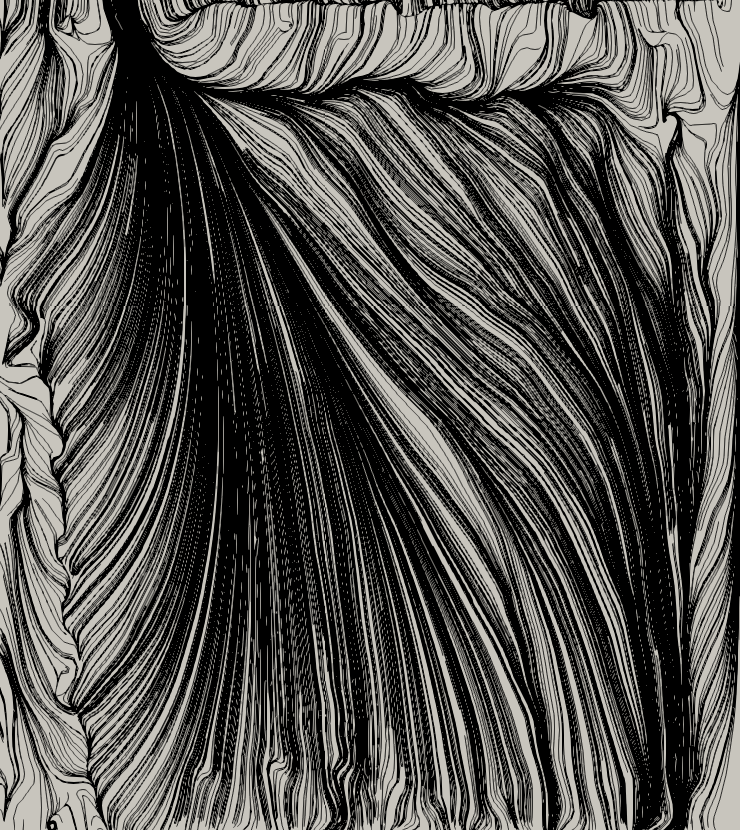}
\caption{DA10}
\label{fig:Ahmed0_wss_DA10}
\end{center}
\end{subfigure}
\begin{subfigure}[b]{0.33\textwidth}
\begin{center}
\includegraphics[width=1\textwidth]{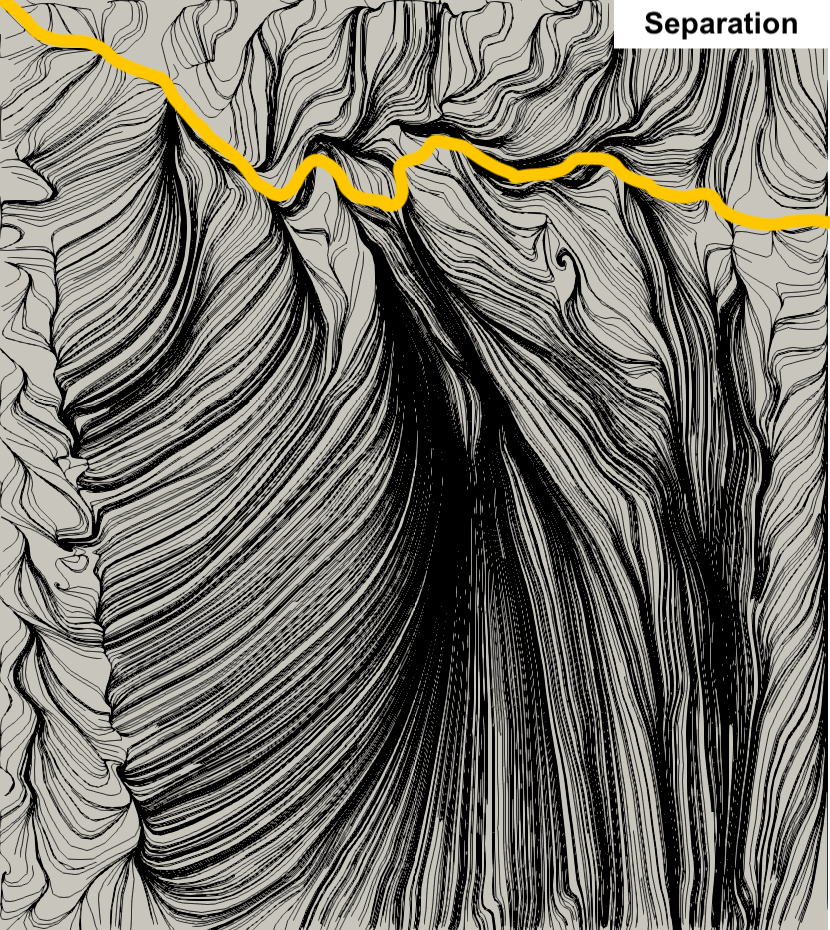}
\caption{DA20}
\label{fig:Ahmed0_wss_DA20}
\end{center}
\end{subfigure}
\begin{subfigure}[b]{0.33\textwidth}
\begin{center}
\includegraphics[width=1\textwidth]{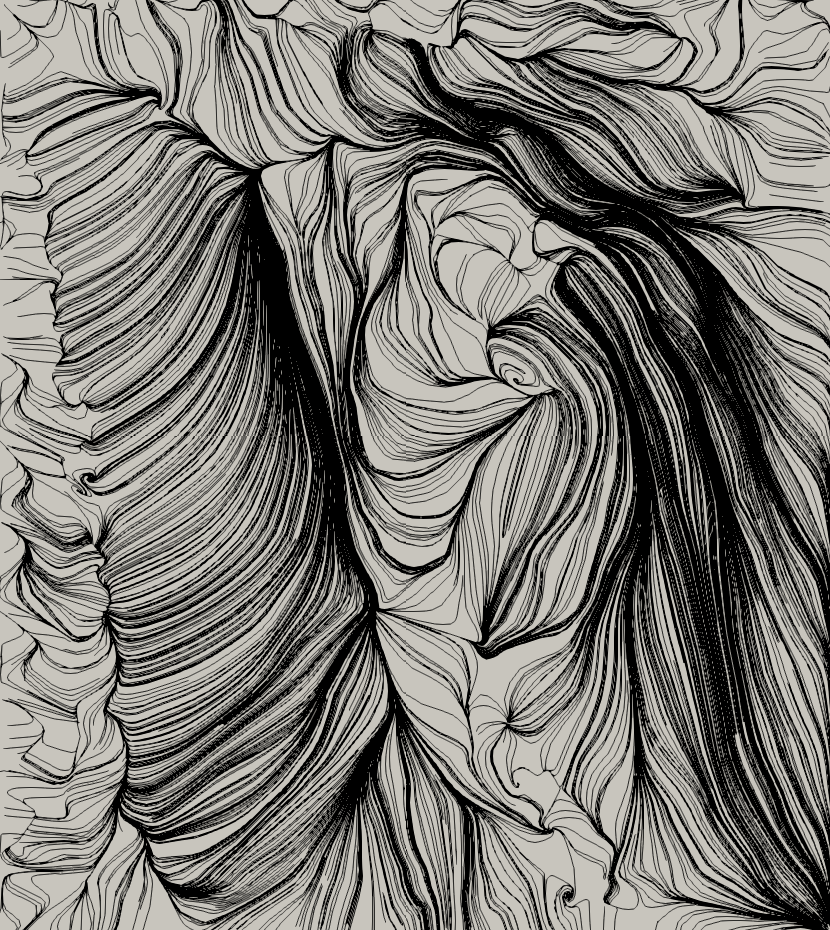}
\caption{DA30}
\label{fig:Ahmed0_wss_DA30}
\end{center}
\end{subfigure}
\begin{subfigure}[b]{0.33\textwidth}
\begin{center}
\includegraphics[width=1\textwidth]{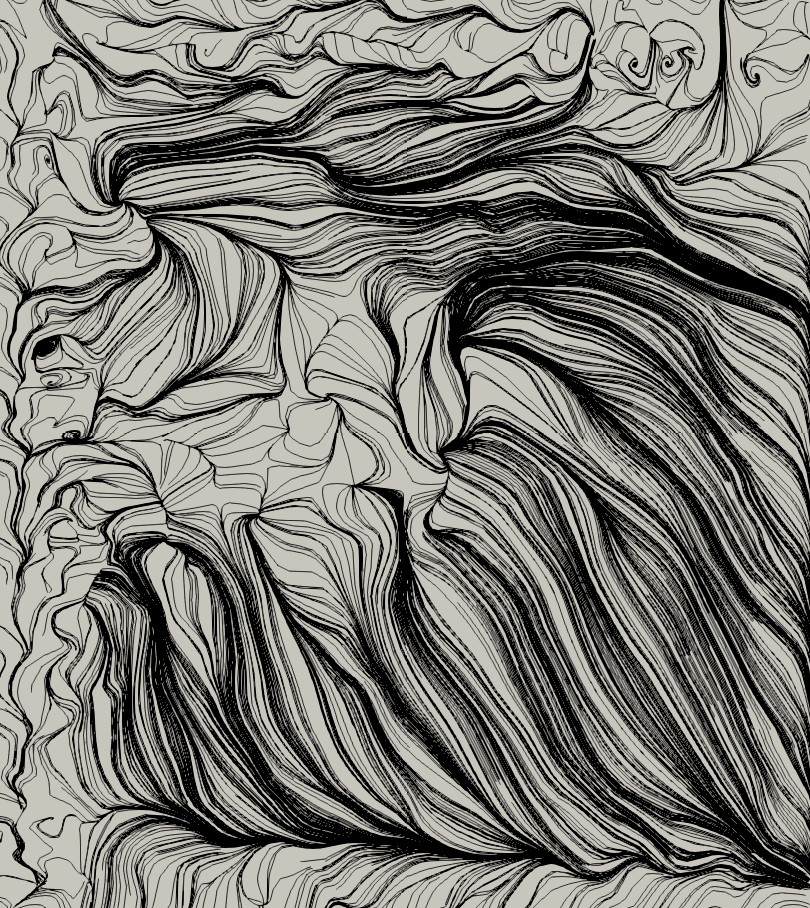}
\caption{DA40}
\label{fig:Ahmed0_wss_DA40}
\end{center}
\end{subfigure}
\begin{subfigure}[b]{0.33\textwidth}
\begin{center}
\includegraphics[width=1\textwidth]{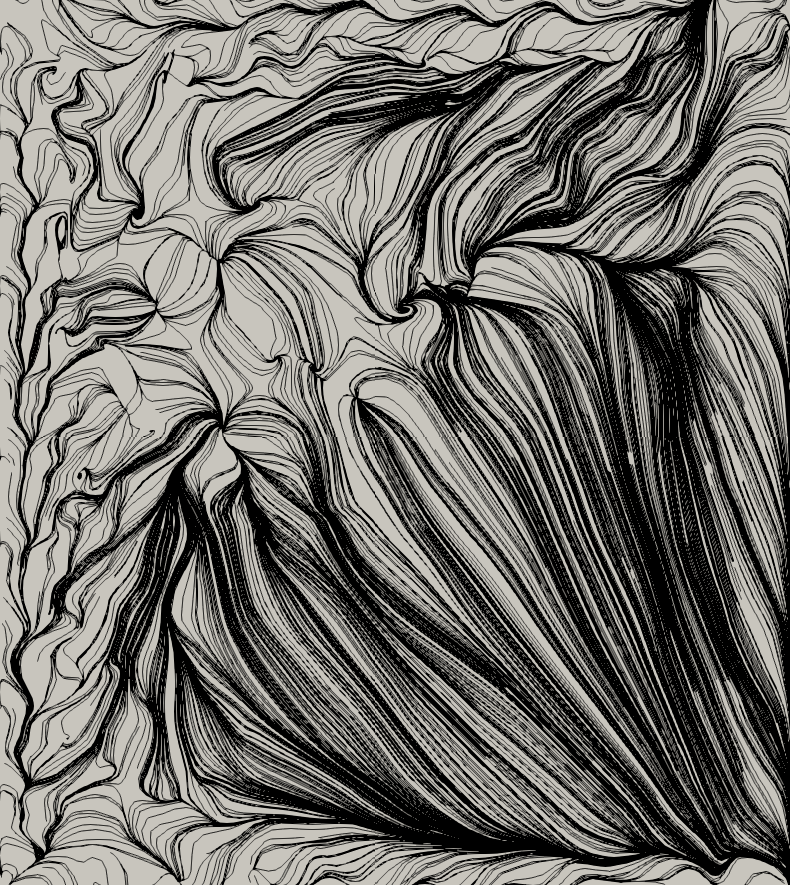}
\caption{DA50}
\label{fig:Ahmed0_wss_DA50}
\end{center}
\end{subfigure}
    \caption{Wall shear stress lines (black) on the diffuser surface for the Ahmed body squared-back considering the proposed diffuser angles: 5$^\circ$ (DA5), 10$^\circ$ (DA10), 20$^\circ$ (DA20), 30$^\circ$ (DA30), 40$^\circ$  (DA40) and 50$^\circ$ (DA50), bottom view, incoming flow direction from top.}
    \label{fig:Ahmed0_wss}
\end{figure}

Averaged wall shear stress lines for the diffuser cases evaluated are next presented. The bottom view of the diffuser, with flow direction coming from the top are shown in Figure~\ref{fig:Ahmed0_wss}. We define the diffuser inlet as the top of the diffuser, and diffuser outlet as the bottom.

Analysing results in Figure~\ref{fig:Ahmed0_wss}, we observe different flow behaviours on the diffuser surface, detailed as follows. The DA5 and DA10 cases have the diffuser vortex influencing the diffuser surface up to the mid-span. There is a clearly defined separation area at the diffuser inlet, responsible for the pumping effect observed by \cite{cooper1998aerodynamic}. The flow keeps attached on the diffuser surface until reaching the diffuser outlet. A combination of both separated flow on the diffuser surface and diffuser vortex is observed for the DA30 case. The diffuser vortex size increases, reaching almost mid-span distance at the diffuser outlet, while fully separated flow is also observed on the diffuser surface. 

The pattern evident in DA20 is a combination of the DA30 and DA10 diffuser flow regimes. There is indication of the diffuser vortex, with defined separation area at the diffuser inlet followed by a small recirculation bubble due to flow separation. The flow then reattaches and follow this pattern until reaching the diffuser outlet. Considering the more extreme diffuser angles (DA40 and DA50), had no previous references, however the low performance is expected. Results show a chaotic behaviour on their surfaces due to the separated flow and a recirculation zone is observed at the diffuser outlet.

\begin{figure}[bt]
\begin{subfigure}[b]{1\textwidth}
\begin{center}
\includegraphics[width=1\textwidth]{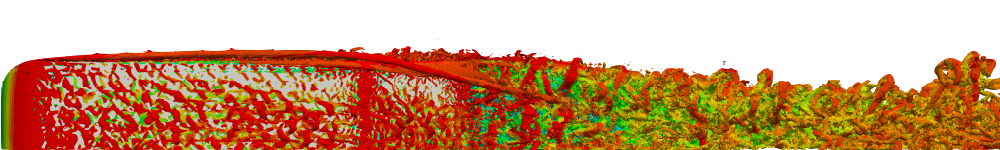}
\caption{DA5}
\label{fig:Ahmed0_BottomU5}
\end{center}
\end{subfigure}
\begin{subfigure}[b]{1\textwidth}
\begin{center}
\includegraphics[width=1\textwidth]{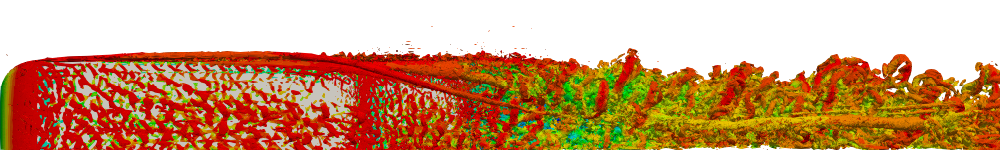}
\caption{DA10}
\label{fig:Ahmed0_BottomU10}
\end{center}
\end{subfigure}
\begin{subfigure}[b]{1\textwidth}
\begin{center}
\includegraphics[width=1\textwidth]{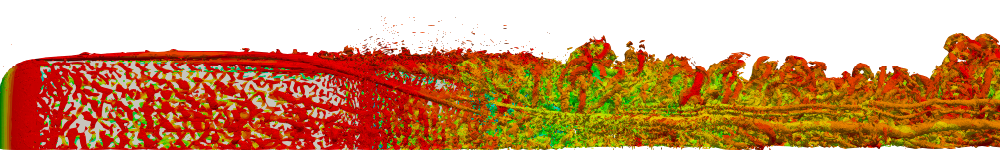}
\caption{DA20}
\label{fig:Ahmed0_BottomU20}
\end{center}
\end{subfigure}
\begin{subfigure}[b]{1\textwidth}
\begin{center}
\includegraphics[width=1\textwidth]{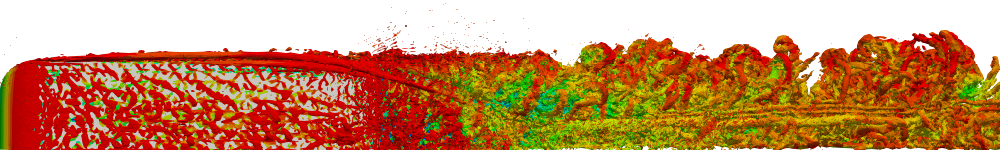}
\caption{DA30}
\label{fig:Ahmed0_BottomU30}
\end{center}
\end{subfigure}
\begin{subfigure}[b]{1\textwidth}
\begin{center}
\includegraphics[width=1\textwidth]{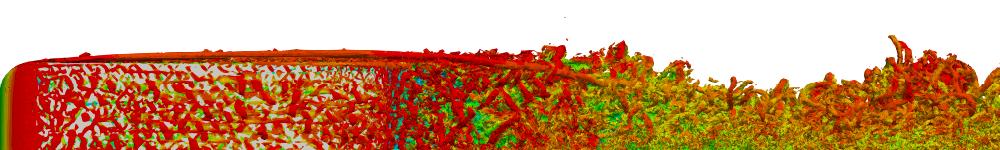}
\caption{DA40}
\label{fig:Ahmed0_BottomU40}
\end{center}
\end{subfigure}
\begin{subfigure}[b]{1\textwidth}
\begin{center}
\includegraphics[width=1\textwidth]{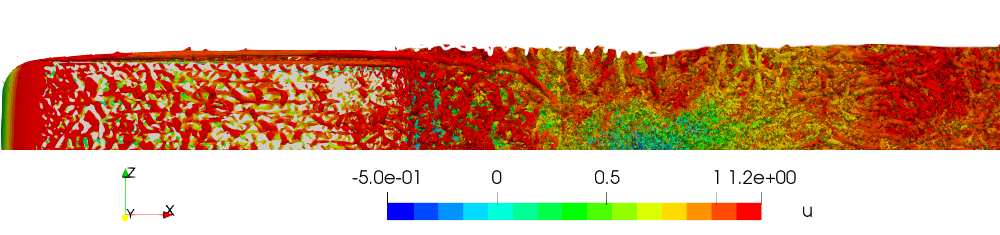}
\caption{DA50}
\label{fig:Ahmed0_BottomU50}
\end{center}
\end{subfigure}
\caption{Iso-contours of Q-Criterion (QCrit = 100), coloured by $U$, of the bottom view of the Ahmed body squared-back. The following proposed diffuser angles are presented: 5$^\circ$ (DA5), 10$^\circ$ (DA10), 20$^\circ$ (DA20), 30$^\circ$ (DA30), 40$^\circ$ (DA40) and 50$^\circ$ (DA50).}
\label{fig:Ahmed0_BottomU}
\end{figure}

Bottom view of the Ahmed body squared-back at different diffuser angles showing iso-contours of Q-Criterion (QCrit = 100) are presented in Figure~\ref{fig:Ahmed0_BottomU}. The lower side vortex is shifting in the inner spanwise direction for DA5 up to DA30.

\subsubsection{Ahmed body 25$^{\circ}$ slant inclination with diffuser}
\label{subsubsec:ahmed25flow}

\begin{figure}[bt]
\begin{subfigure}[b]{1\textwidth}
\begin{center}
\makebox[\textwidth][c]{\includegraphics[width=1.3\textwidth]{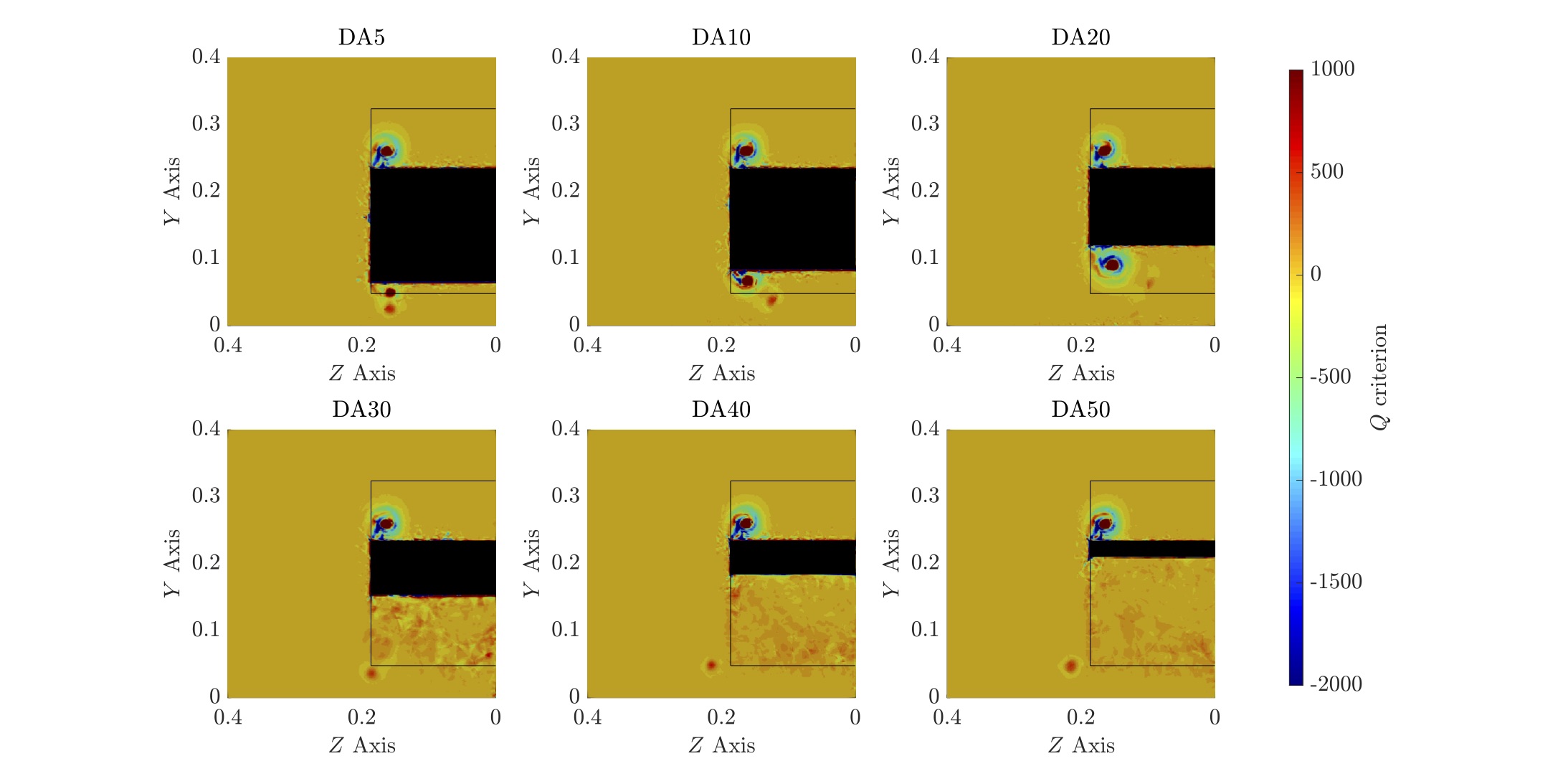}}%
\caption{Plane $X/L =0$}
\label{fig:Ahmed25_Plane0_Q}
\end{center}
\end{subfigure}
\begin{subfigure}[b]{1\textwidth}
\begin{center}
\makebox[\textwidth][c]{\includegraphics[width=1.3\textwidth]{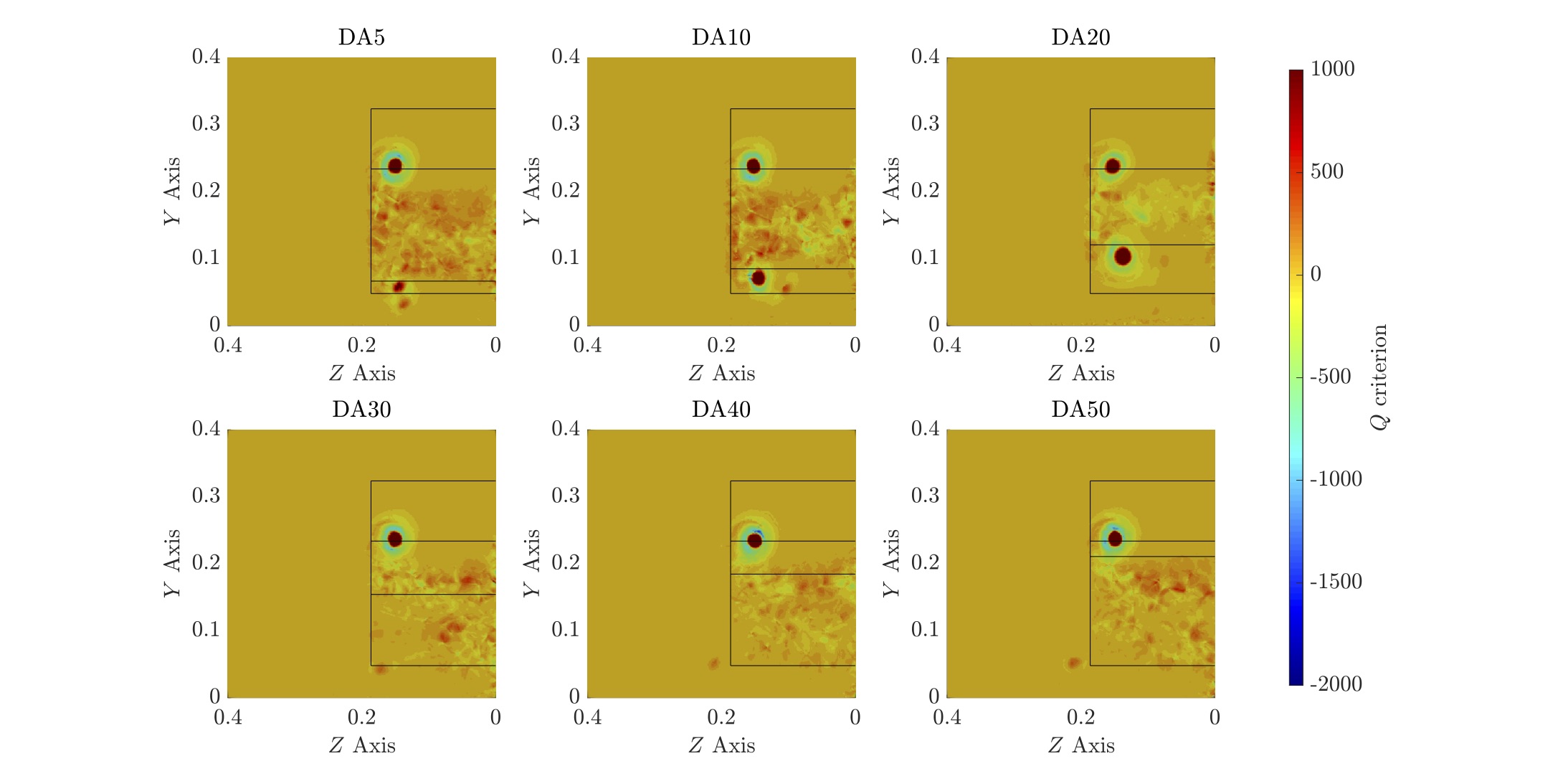}}%
\caption{Plane $X/L =0.096$}
\label{fig:Ahmed25_Plane096_Q}
\end{center}
\end{subfigure}
    \caption{Contours of Q-Criterion (QCrit = 100) for the Ahmed body with slant angle of 25$^\circ$ considering diffuser angle of 5$^\circ$ (DA5), 10$^\circ$ (DA10), 20$^\circ$ (DA20), 30$^\circ$ (DA30), 40$^\circ$  (DA40) and 50$^\circ$ (DA50) for planes $X/L =0$ and $X/L =0.096$.}
\label{fig:Ahmed25_Q}
\end{figure}

We now analyse flow structures found for the Ahmed body with slant angle of 25$^{\circ}$ equipped with underbody diffuser. On the upper part of the body, the slant vortex is clearly defined in all diffuser angles evaluated. As presented for previous case, contours of Q-Criterion for planes $X/L = 0$ and $X/L = 0.096$ are shown in Figure~\ref{fig:Ahmed25_Q}. The diffuser vortex appears on DA5, DA10 and DA20 together with an additional small intensity vortex on the plane $X/L = 0$. The diffuser vortex has similar intensity and size as the slant vortex in the DA20 case. The other three diffuser cases, DA30, DA40 and DA50 indicate no evidence of the diffuser vortex but only the same lower side vortex, originated on the frontal part of the Ahmed body. Further downstream on plane $X/L = 0.096$, the vortical system have moved inward in the spanwise direction on the first three cases. The last three cases indicate that the lower side vortex gets weaker as it moved downstream.

\begin{figure}[bt]
\begin{subfigure}[b]{1\textwidth}
\begin{center}
\makebox[\textwidth][c]{\includegraphics[width=1.3\textwidth]{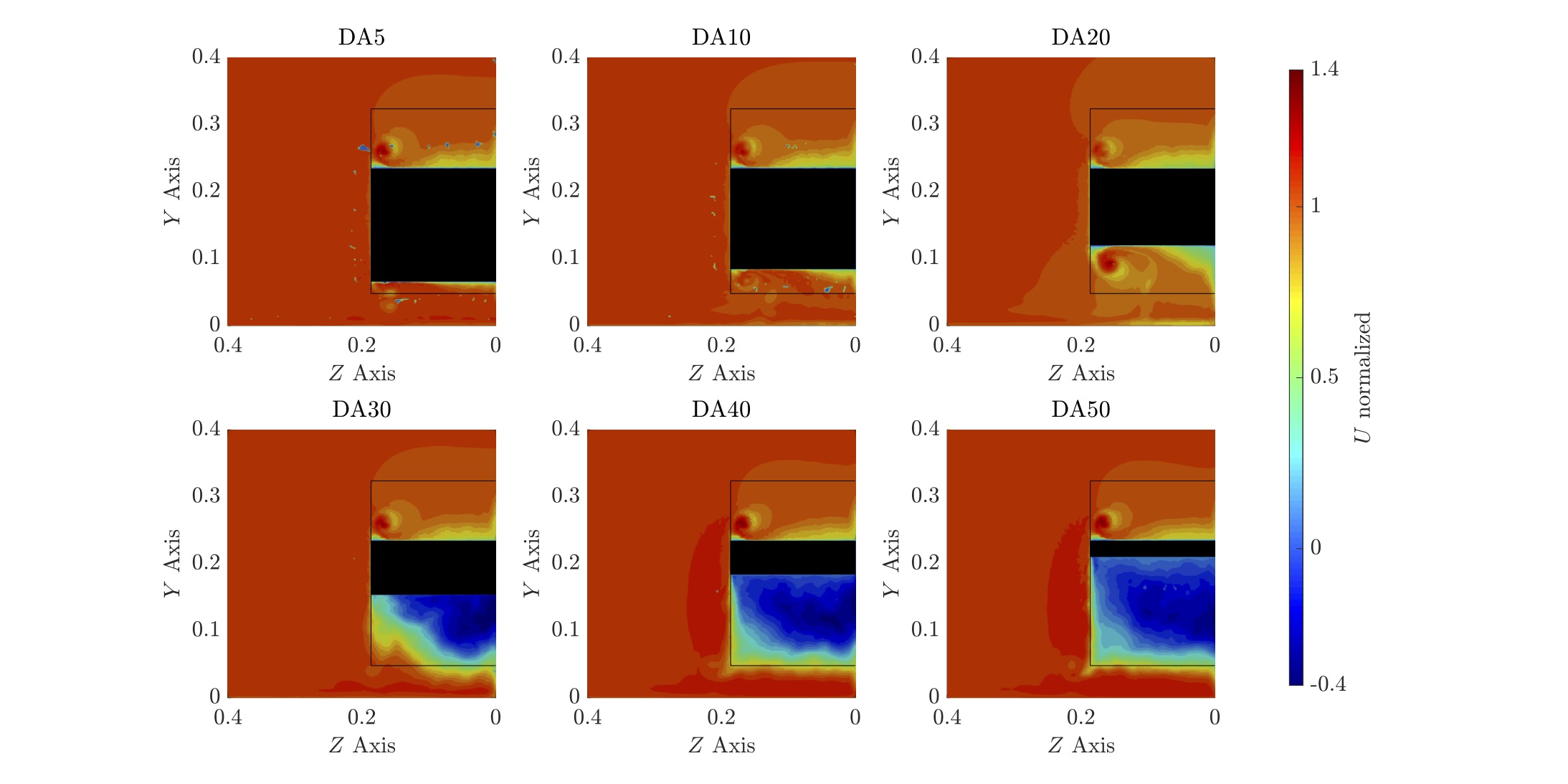}}%
\caption{Plane $X/L =0$}
\label{fig:Ahmed25_Plane0_U}
\end{center}
\end{subfigure}
\begin{subfigure}[b]{1\textwidth}
\begin{center}
\makebox[\textwidth][c]{\includegraphics[width=1.3\textwidth]{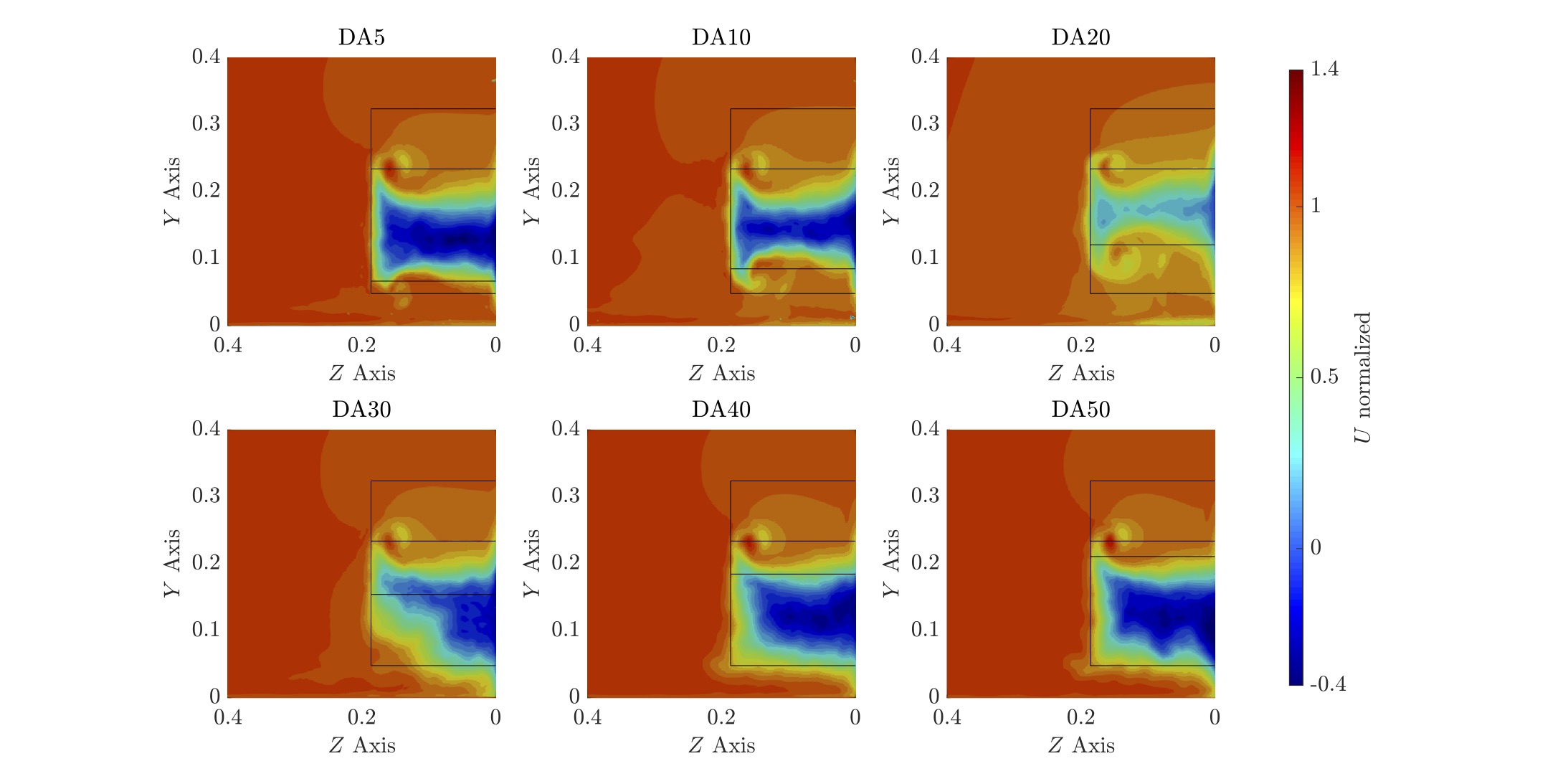}}%
\caption{Plane $X/L =0.096$}
\label{fig:Ahmed25_Plane096_U}
\end{center}
\end{subfigure}
    \caption{Contours of normalized streamwise velocity $U$ for the Ahmed body with slant angle of 25$^\circ$ considering the proposed diffuser angles: 5$^\circ$ (DA5), 10$^\circ$ (DA10), 20$^\circ$ (DA20), 30$^\circ$ (DA30), 40$^\circ$  (DA40) and 50$^\circ$ (DA50) for planes $X/L =0$ and $X/L =0.096$.}
\label{fig:Ahmed25_U}
\end{figure}

Contours of normalized $U$ for both $X/L = 0$ and $X/L =0.096$ planes are presented in Figure~\ref{fig:Ahmed25_U}. On plane $X/L = 0$, we observe that the velocity countour on the upper slant changes as the diffuser angle becomes more inclined until the case DA20. The three other cases from DA30 to DA50 however, have similar velocity profiles. We conclude that the diffuser influences the flow over the upper slant whenever we have evidences of the diffuser vortex. From observation of the first Ahmed body case, we have indication of attached flow for diffuser angle up to 20$^\circ$, whereas for higher angles, a significant wake contribution can be seen from the fully separated flow from the diffuser. 

Moving downstream, normalized $U$ velocity on plane $X/L = 0.096$ shows the evolution of the turbulent wake and vortices. Base pressure turbulent wake with the slant and diffuser vortices are the main structures seen on cases DA5 and DA10. The $U$ velocity contours on this plane for the DA20 case shows a very small negative velocity zone, indicating an energetic wake, together with the slant and diffuser vortices moving downstream. On DA30 case, wake profile and vortical system is similar to DA40 and DA50, except by the fact of a distortion on the lower outer area of the diffuser. At this point, the flow distortion could be caused by a vortical flow structure and further plots will provide evidences to confirm this assumption.

\begin{figure}[bt]
\begin{subfigure}[b]{1\textwidth}
\begin{center}
\makebox[\textwidth][c]{\includegraphics[width=1.3\textwidth]{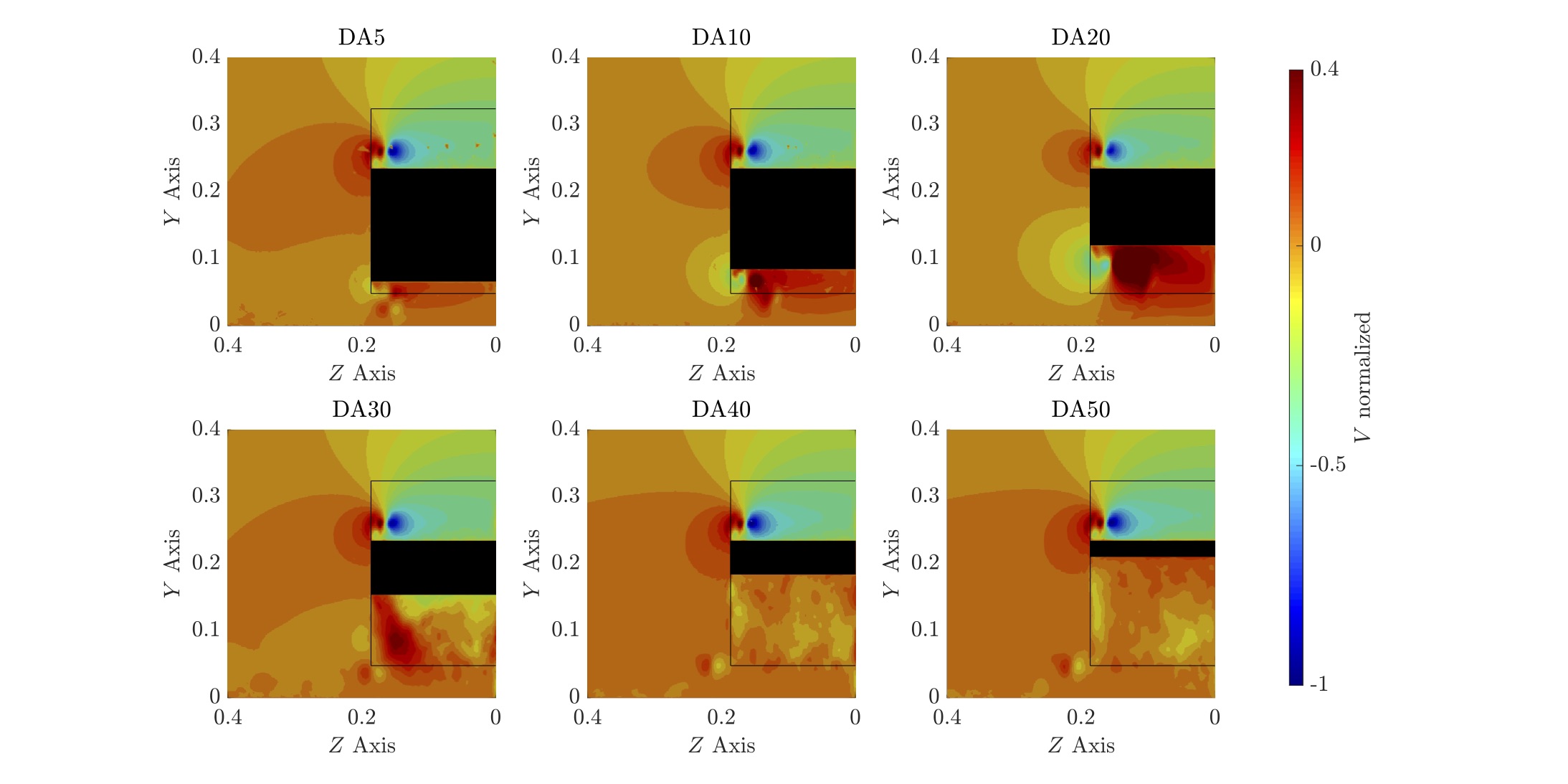}}%
\caption{Plane $X/L =0$}
\label{fig:Ahmed25_Plane0_V}
\end{center}
\end{subfigure}
\begin{subfigure}[b]{1\textwidth}
\begin{center}
\makebox[\textwidth][c]{\includegraphics[width=1.3\textwidth]{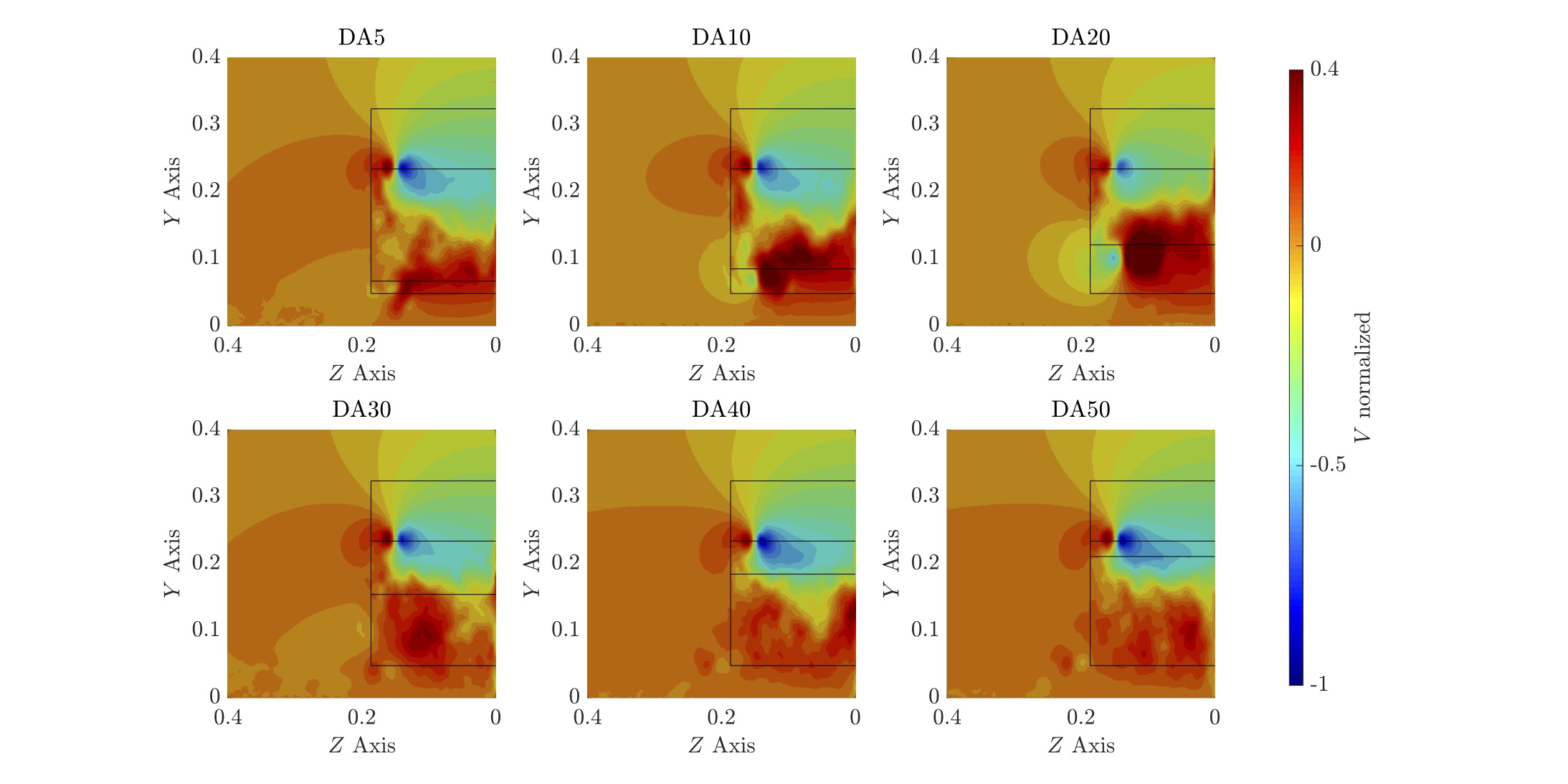}}%
\caption{Plane $X/L =0.096$}
\label{fig:Ahmed25_Plane096_V}
\end{center}
\end{subfigure}
    \caption{Contours of normalized vertical velocity $V$ for the Ahmed body with slant angle of 25$^\circ$ considering the proposed diffuser angles: 5$^\circ$ (DA5), 10$^\circ$ (DA10), 20$^\circ$ (DA20), 30$^\circ$ (DA30), 40$^\circ$ (DA40) and 50$^\circ$ (DA50) for planes $X/L =0$ and $X/L =0.096$.}
\label{fig:Ahmed25_V}
\end{figure}

Normalized vertical velocity $V$ is presented in Figure~\ref{fig:Ahmed25_V} where we observe similar contour on the slant as presented by the experimental reference of \cite{lienhart2003flow} on plane $X/L = 0$. When analysing the diffuser area, two vortices are identified at similar spanwise coordinates but different heights on DA5 case. From bottom to top, the lower side vortex and the diffuser vortex are in the same region, however it is not possible to assure they are merging at this point. Only the diffuser vortex is observed for DA10 and DA20 with anti-clockwise rotation direction. The case considering diffuser angle at 30$^\circ$ (DA30) also has an indication of two vortices, however only the lower side vortex (bottom) can be confirmed at this point. For the DA40 and DA50 cases, the lower side vortex has similar intensity in both case and a slightly different $V$ velocity distribution on the diffuser, with higher velocities in the most inclined diffuser.

Analysing plane $X/L = 0.096$, the inner spanwise component of the slant vortex is shifting downwards, once this flow structure starts to interact with the wake. On the diffuser area of DA5, the pair of vortices is merged into one single structure, with high $V$ velocity on the positive component of the new merged vortex. DA10 and DA20 cases maintain only one single diffuser vortex structure highlighted, moving slightly up and into the spanwise direction. The DA30 case still maintain the lower side vortex in similar position as in plane $X/L = 0$ and the additional structure does not behave as a vortex. The diffuser wake structure is still similar on DA40 compared to DA50, where the main difference relies on a low velocity zone at the mid-span of the diffuser, caused by the interference of slant vortex on the diffuser wake. 

\begin{figure}[hbt]
\begin{subfigure}[b]{0.33\textwidth}
\begin{center}
\includegraphics[width=1\textwidth]{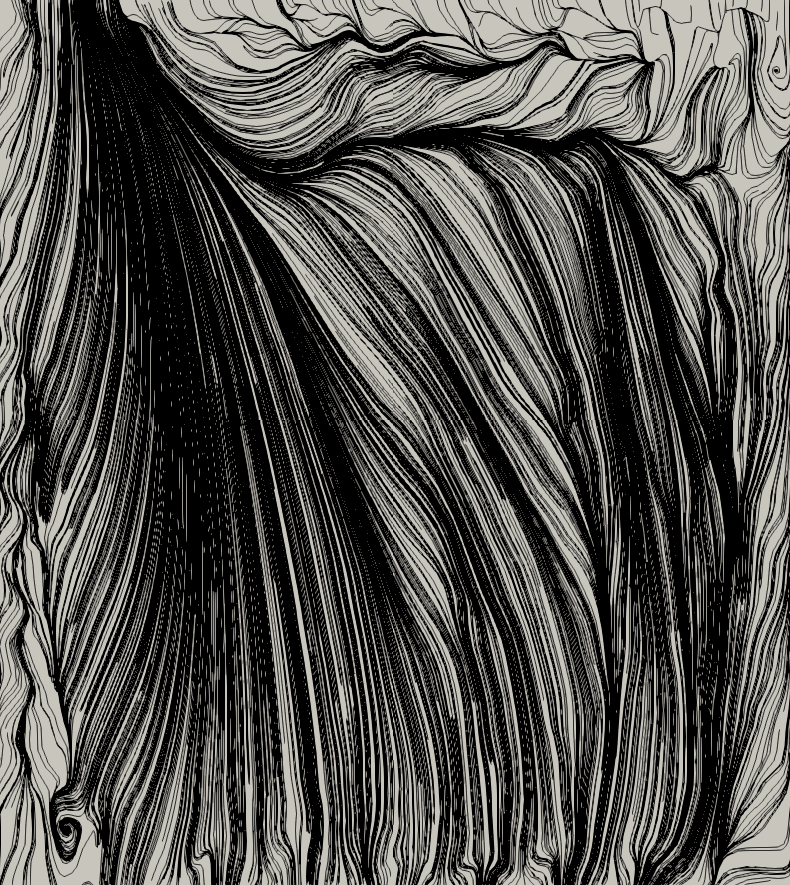}
\caption{DA5}
\label{fig:Ahmed25_wss_DA5}
\end{center}
\end{subfigure}
\begin{subfigure}[b]{0.33\textwidth}
\begin{center}
\includegraphics[width=1\textwidth]{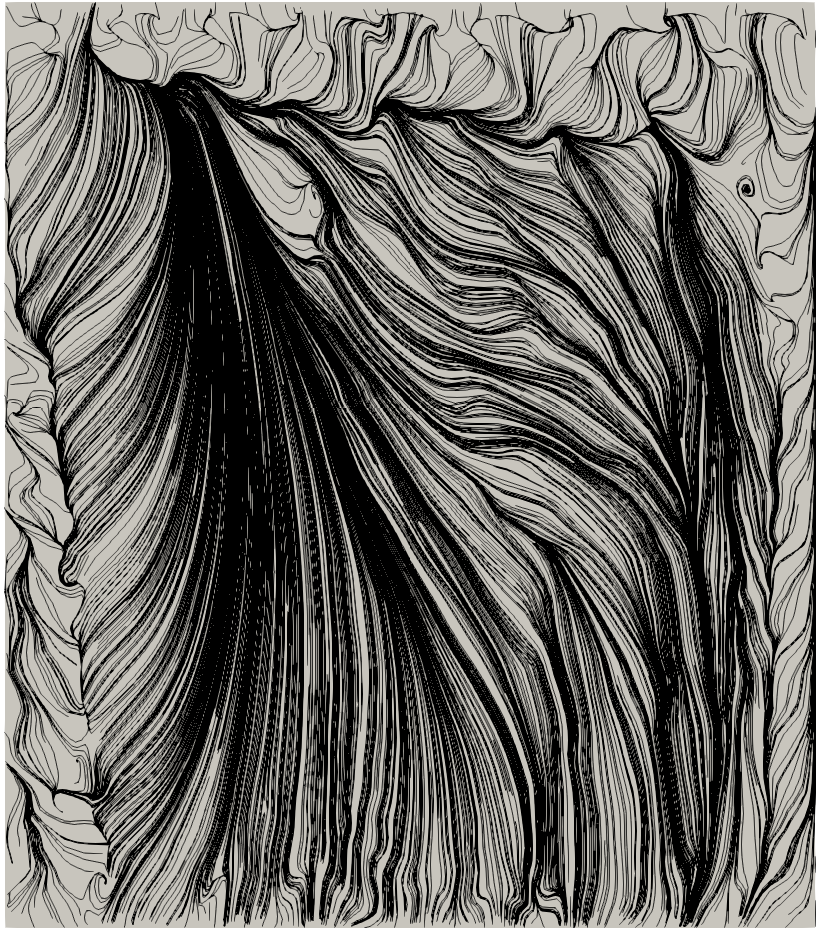}
\caption{DA10}
\label{fig:Ahmed25_wss_DA10}
\end{center}
\end{subfigure}
\begin{subfigure}[b]{0.33\textwidth}
\begin{center}
\includegraphics[width=1\textwidth]{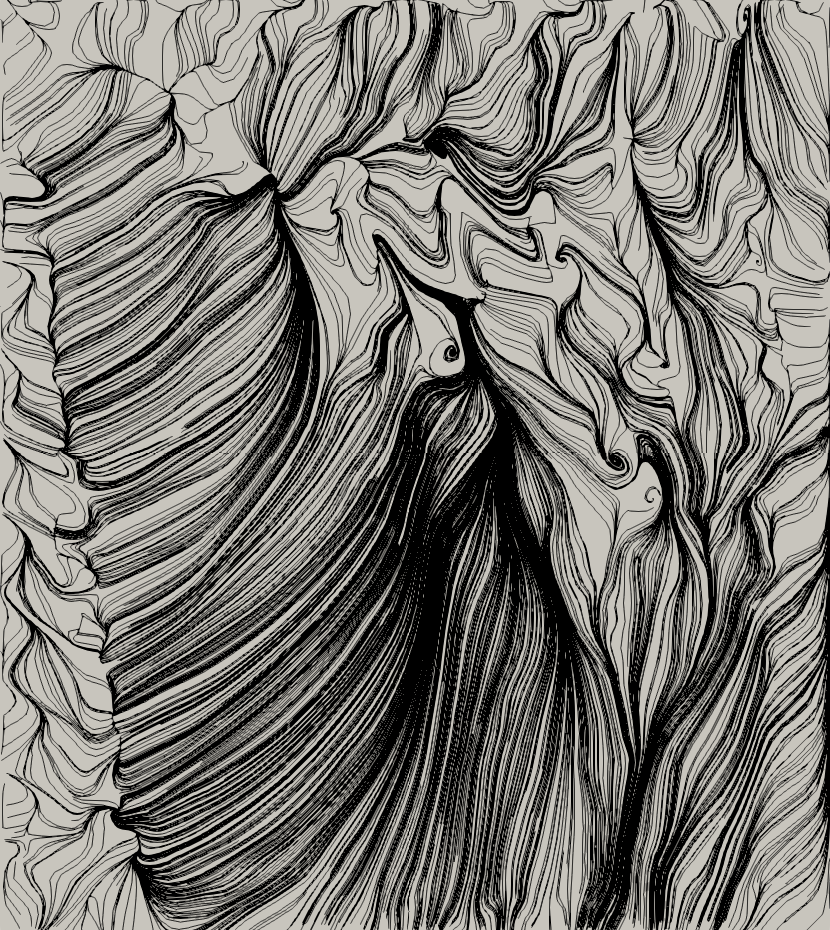}
\caption{DA20}
\label{fig:Ahmed25_wss_DA20}
\end{center}
\end{subfigure}
\begin{subfigure}[b]{0.33\textwidth}
\begin{center}
\includegraphics[width=1\textwidth]{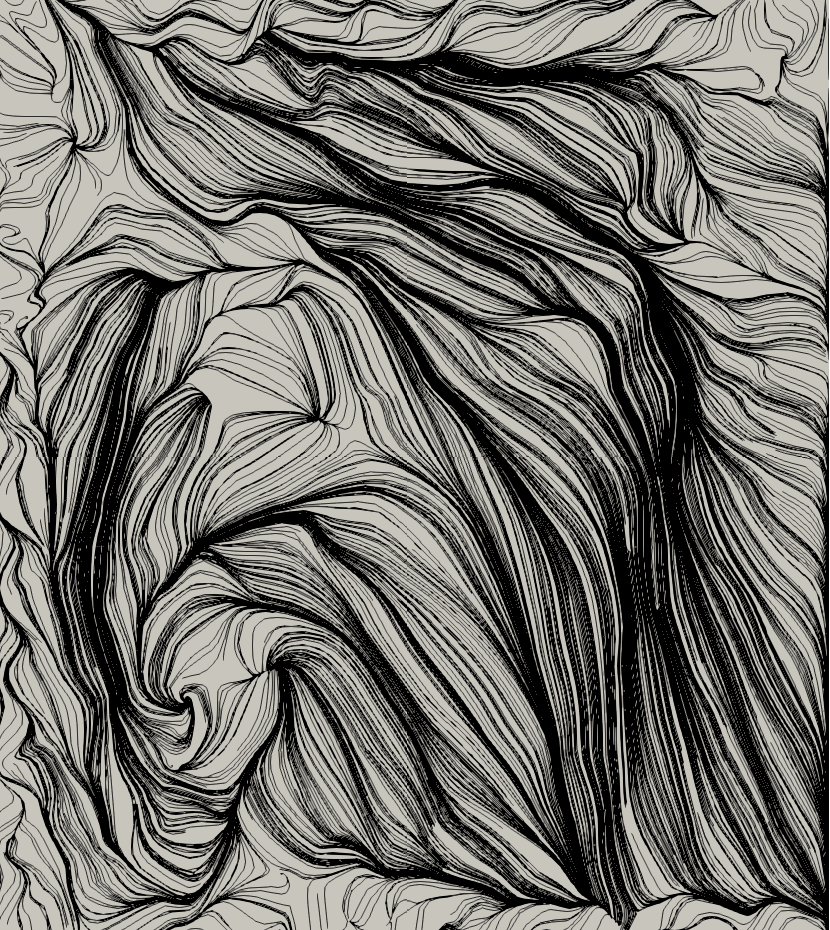}
\caption{DA30}
\label{fig:Ahmed25_wss_DA30}
\end{center}
\end{subfigure}
\begin{subfigure}[b]{0.33\textwidth}
\begin{center}
\includegraphics[width=1\textwidth]{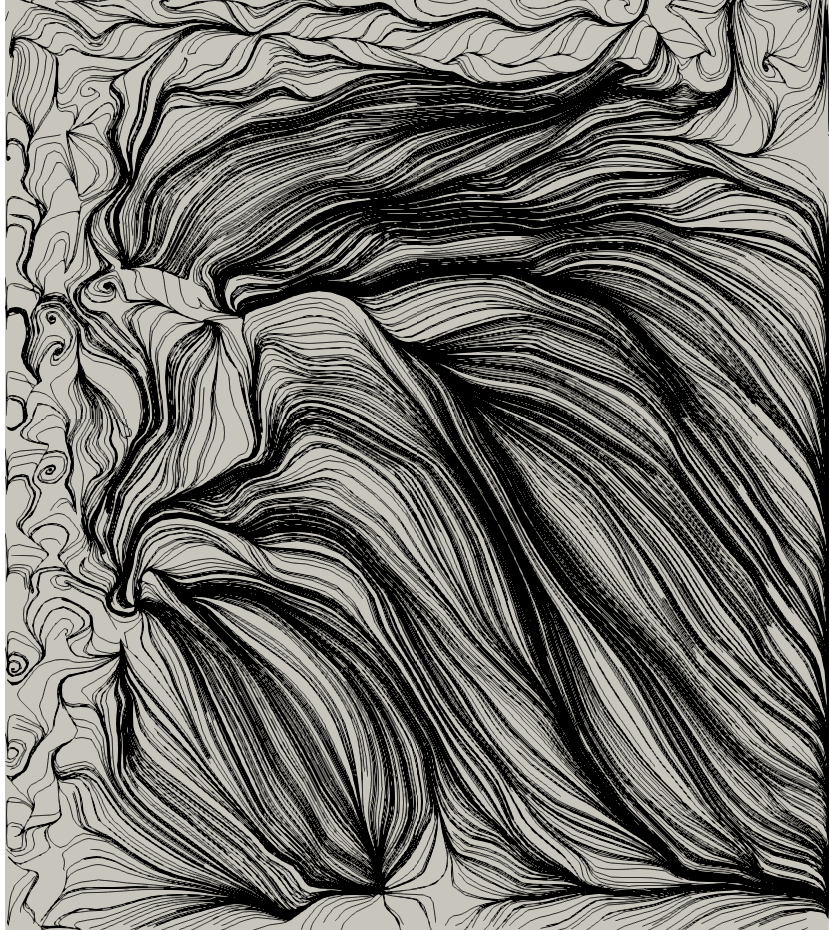}
\caption{DA40}
\label{fig:Ahmed25_wss_DA40}
\end{center}
\end{subfigure}
\begin{subfigure}[b]{0.33\textwidth}
\begin{center}
\includegraphics[width=1\textwidth]{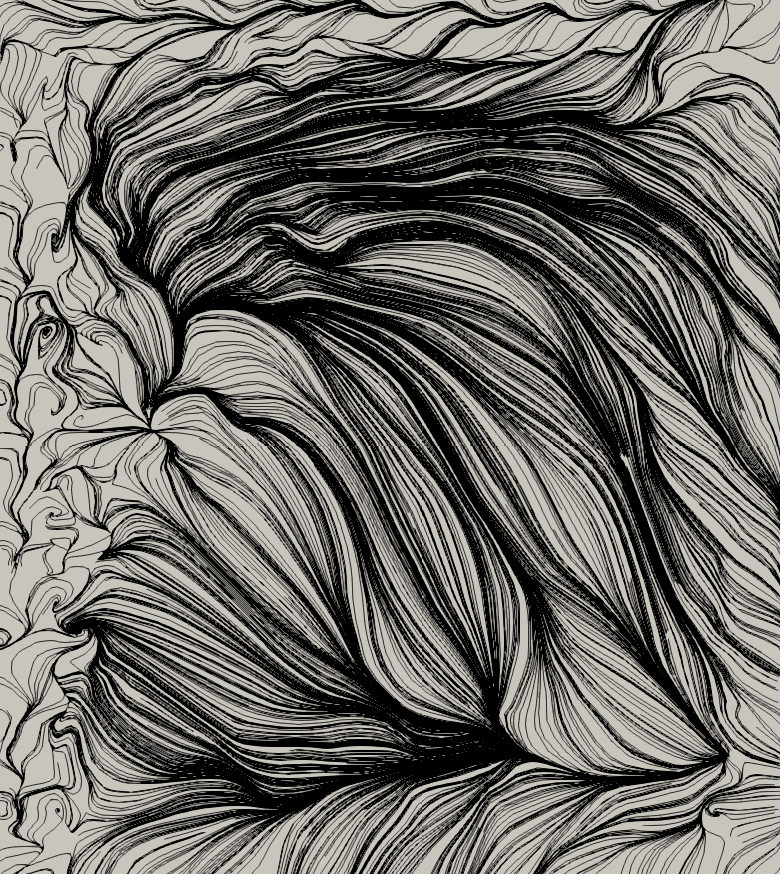}
\caption{DA50}
\label{fig:Ahmed25_wss_DA50}
\end{center}
\end{subfigure}
    \caption{Wall shear stress lines (black) on the diffuser surface for the Ahmed body with slant angle of 25$^\circ$ considering the proposed diffuser angles: 5$^\circ$ (DA5), 10$^\circ$ (DA10), 20$^\circ$ (DA20), 30$^\circ$ (DA30), 40$^\circ$  (DA40) and 50$^\circ$ (DA50), bottom view, incoming flow direction from top.}
    \label{fig:Ahmed25_wss}
\end{figure}

Wall shear stress lines on the surface of each diffuser case evaluated are presented in Figure~\ref{fig:Ahmed25_wss}. This analysis follows similar terminology and setup as presented for the Ahmed body squared-back.

The flow structure on the diffuser indicates a vortex touching the diffuser surface up to angle of 20$^\circ$ (DA20) and separated flow from DA30 onward. We identified three flow behaviour on the diffuser surface as the previous Ahmed body squared-back, detailed as follows. The DA5 and DA10 cases have the side diffuser vortex touching the diffuser surface together with a separation area at the diffuser inlet. The flow remains attached on the surface until reaching the outlet. Separated flow on the diffuser surface and diffuser vortex is observed in the DA20 case, with a partial reattachment at the outlet. The last three diffuser angles of 30$^\circ$ (DA30), 40$^\circ$ (DA40) and 50$^\circ$ (DA50) have a fully separated flow all over the diffuser surface. 

\begin{figure}[bt]
\begin{subfigure}[b]{1\textwidth}
\begin{center}
\includegraphics[width=1\textwidth]{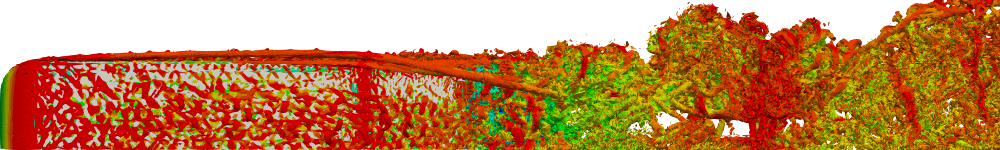}
\caption{DA5}
\label{fig:Ahmed25_BottomU5}
\end{center}
\end{subfigure}
\begin{subfigure}[b]{1\textwidth}
\begin{center}
\includegraphics[width=1\textwidth]{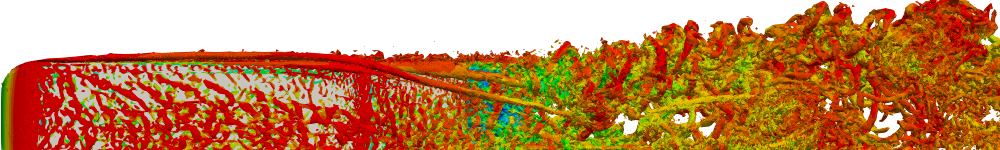}
\caption{DA10}
\label{fig:Ahmed25_BottomU10}
\end{center}
\end{subfigure}
\begin{subfigure}[b]{1\textwidth}
\begin{center}
\includegraphics[width=1\textwidth]{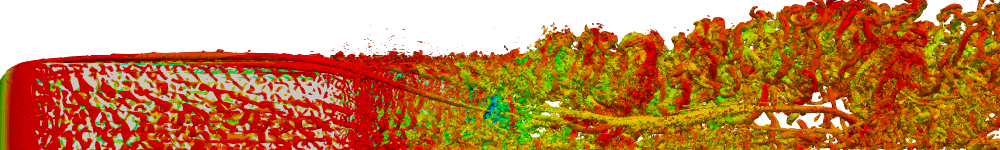}
\caption{DA20}
\label{fig:Ahmed25_BottomU20}
\end{center}
\end{subfigure}
\begin{subfigure}[b]{1\textwidth}
\begin{center}
\includegraphics[width=1\textwidth]{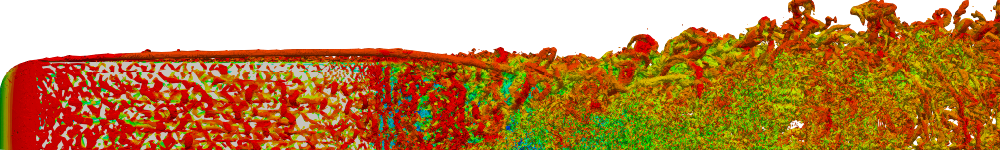}
\caption{DA30}
\label{fig:Ahmed25_BottomU30}
\end{center}
\end{subfigure}
\begin{subfigure}[b]{1\textwidth}
\begin{center}
\includegraphics[width=1\textwidth]{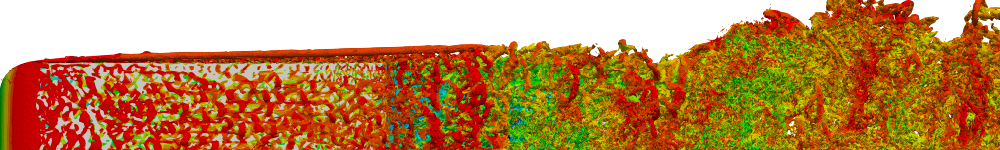}
\caption{DA40}
\label{fig:Ahmed25_BottomU40}
\end{center}
\end{subfigure}
\begin{subfigure}[b]{1\textwidth}
\begin{center}
\includegraphics[width=1\textwidth]{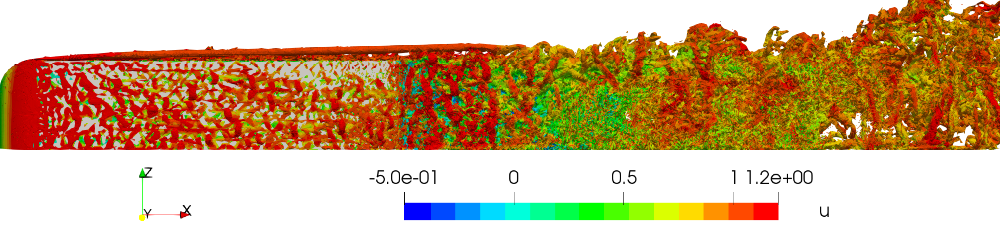}
\caption{DA50}
\label{fig:Ahmed25_BottomU50}
\end{center}
\end{subfigure}
\caption{Iso-contours of Q-Criterion (QCrit = 100), coloured by $U$, of the bottom view of the Ahmed body with slant angle of 25$^\circ$. The following proposed diffuser angles are presented: 5$^\circ$ (DA5), 10$^\circ$ (DA10), 20$^\circ$ (DA20), 30$^\circ$ (DA30), 40$^\circ$ (DA40) and 50$^\circ$ (DA50).}
\label{fig:Ahmed25_BottomU}
\end{figure}

Flow structures on the bottom view for the Ahmed body with slant angle of 25$^\circ$ considering different diffuser angles are presented in Figure~\ref{fig:Ahmed25_BottomU}. From the iso-contours of Q-Criterion (QCrit = 100) coloured by $U$, we observe the same lower side vortex behaviour shifting inwards in the spanwise direction. The lower vortex shifting only happens with the existence of the diffuser vortex, from DA5 to DA20 in this case.

\section{Conclusions}
\label{sec:6}

A parametric study on the effect of diffusers is conducted on the Ahmed body at two slant cases: squared-back and 25$^{\circ}$ angle. The diffuser length is fixed with the same dimension of the slant and cases are evaluated at a Reynolds number of $Re = 1.7 \times 10^6$ with moving ground condition. Diffuser angles considered range from 10$^{\circ}$ to 50$^{\circ}$ in increments of 10$^{\circ}$, including one additional angle of 5$^{\circ}$ for both cases. The numerical methodology employed in this study was validated on the classical Ahmed body geometry, where it was found good agreement in terms of flow structures and the difference of 13\% and 1\% for respectively drag and lift coefficient against experiments. The application of the spectral/hp element method also allowed to have a high-fidelity solution from a relatively coarse mesh. Further increments on the resolution by increasing the solution polynomial order $P_N$ are possible, keeping the same mesh structure, which is highly desirable for comparative purposes. The study extended the knowledge on plane diffusers by showing the flow behaviour on a well-established bluff body when the downforce generation by the diffuser is not effective anymore. The interaction of the lower side vortices with the flow structures on the diffuser region due to the absence of endplates is an interesting phenomenon.

Ahmed body squared-back results indicate two different flow behaviours. The first is observed in diffuser angles from 5$^{\circ}$ up to 30$^{\circ}$, indicating that downforce increment leads to higher drag coefficient. Flow structure for this regime is composed of a lateral vortex with fully attached flow on the diffuser surface for diffusers up to 20$^{\circ}$. The critical angle has similar structure considering the same diffuser vortex however the flow is partially separated on the diffuser surface. The second flow regime is found for diffuser angles higher than 30$^{\circ}$ where downforce increment efficiency is lost, and the flow is fully separated on the diffuser surface.

On the Ahmed body with 25$^{\circ}$ slant angle, downforce increases while the drag coefficient is reduced for diffuser angles up to 20$^{\circ}$. Maximum downforce is observed at 10$^{\circ}$. The flow structure is composed of the diffuser vortex and attached flow on the diffuser surface, however only for the 5$^{\circ}$ and 10$^{\circ}$ diffuser angles. The diffuser vortex is present in the 20$^{\circ}$ case but the flow is mostly separated. Similar to the squared-back case, the highest drag coefficient value is observed for the 30$^{\circ}$ angle, and flow on the diffuser surface is fully separated from this case onward.

The downforce enhancement provided by the diffuser is connected to the presence of the diffuser vortex on the lower portion of the body. As the diffuser vortex disappear and the flow becomes fully separated, only small downforce increments are observed. Before the diffuser flow separates, the effect of an increment in the diffuser angle aims to increase the intensity of the diffuser vortex, which is associated with an increase in downforce.  We therefore conclude that an efficient diffuser needs to  have a coherent diffuser vortex present. 

The flow topology on the diffuser surface is also modified by the diffuser angle changes, from attached flow to separated flow both in the presence of the diffuser vortex. At this point, the most efficient diffuser angle in terms of downforce presents diffuser vortex and separated flow as its flow structure.

The diffuser is a passive device that is influenced by the base-pressure region. Results of the aerodynamic quantities show that although following similar trends, the downforce increment is higher for the Ahmed body squared-back, which has higher base-pressure region and also offsets by 10$^{\circ}$, the most efficiency diffuser angle when compared to the Ahmed body with 25$^{\circ}$ slant angle. 

The drag coefficient follows the opposite trends for each Ahmed body due to different flow structures on the upper slant. Considering the cases where the diffuser is most efficient, and the flow on the diffuser surface changes from attached to separated, there is a drag increment. This drag increment is very small for the Ahmed body with 25$^{\circ}$ slant angle but approximately 45\% for the Ahmed body squared-back for a downforce increment of approximately 7\%. 

We conclude that the diffuser performance is connected to the body geometry and the highest downforce increment is reached once the flow structure is composed by the diffuser vortex and separated flow. For the overall aerodynamic performance, the drag coefficient should be considered in order to select the best diffuser angle.

\begin{acknowledgement}
We acknowledge the HPC facilities at Imperial College Research Computing Service,  DOI: 10.14469/hpc/2232 and also under the UK Turbulence Consortium. CNPq for the sponsorship 202578/2015-1 and EPSRC under the grant.
\end{acknowledgement}

\bibliographystyle{plainnat}
\bibliography{Ahmed_Diffuser_0.bib}

\end{document}